\documentclass[fleqn,usenatbib,useAMS]{mnras}

\usepackage[T1]{fontenc}

\DeclareRobustCommand{\VAN}[3]{#2}
\let\VANthebibliography\thebibliography
\def\thebibliography{\DeclareRobustCommand{\VAN}[3]{##3}\VANthebibliography}

\usepackage{hyperref}
\usepackage{subcaption}
\usepackage[utf8]{inputenc}
\usepackage{graphicx}	
\usepackage{amsmath}	
\usepackage{amssymb}	
\usepackage{tablefootnote}

\def\mass{$M_{\rm{\odot}}$}
\def\kms{km~s$^{-1}$}
\def\siii{\ion{Si}{ii}~$\lambda$6\,355}
\def\mgiin{\ion{Mg}{ii}~$\lambda$9\,227}
\def\mgiit{\ion{Mg}{ii}~$\lambda$10\,952}
\def\feiit{\ion{Fe}{ii}~$\lambda$6\,247}
\def\feiif{\ion{Fe}{ii}~$\lambda$6\,456}

\def\phn{\phantom{0}}

\usepackage{newtxtext,newtxmath}

\title[Mixing in SNe~Iax]{An analysis of the spectroscopic signatures of layering in the ejecta of type~Iax supernovae}

\author[M. R. Magee et al.]{
M. R. Magee$^{1,2}$\thanks{E-mail: mrmagee.astro@gmail.com},
J. H. Gillanders$^2$,
K. Maguire$^1$,
S. A. Sim$^2$,
F. P. Callan$^2$
\\
$^{1}$School of Physics, Trinity College Dublin, The University of Dublin, Dublin 2, Ireland \\ 
$^{2}$Astrophysics Research Centre, School of Mathematics and Physics, Queen's University Belfast, Belfast, BT7 1NN, UK\\                                                   
}

\date{Accepted 2021 October 22. Received 2021 August 12; in original form 2020 November 20}

\pubyear{2021}

\begin{document}
\label{firstpage}
\pagerange{\pageref{firstpage}--\pageref{lastpage}}
\maketitle

\begin{abstract}
Investigations of some type Iax supernovae have led to the suggestion that their ejecta must be layered to some degree. Such an ejecta structure has been argued as inconsistent with the well-mixed composition predicted by pure deflagrations. Based on explosion models, we create toy models in which the ejecta are artificially stratified and progressively mixed until a uniform composition is obtained. We find that models that are heavily mixed, containing burned and unburned material at all velocities, produce reasonably good agreement with SN 2012Z, for which a layered structure has been suggested. We also discuss how existing ejecta compositions determined for type Iax supernovae do not necessarily contradict pure deflagration models and may be consistent with a steeper density profile. We investigate previous claims that differences in line profile shapes may be due to strong blending, by presenting a series of models with different plasma states. These models indicate that blending could indeed explain differences in the observed profiles. Alternatively, stratification could also explain such differences, however all of our models indicate that this does not necessarily require stratification in abundance. Sufficient stratification in ionisation state can be achieved even for a well-mixed model. Based on our analysis, we demonstrate that there is insufficient evidence to suggest the ejecta of type Iax supernovae must be layered and therefore argue the pure deflagration scenario is not ruled out, even for the brightest type Iax supernovae. Our analysis does not indicate the ejecta cannot be layered to some degree, but observations within days of explosion are necessary to determine the extent to which the outer ejecta could be layered.

\end{abstract}




\section{Introduction}
\label{sect:intro}

The thermonuclear explosion of a white dwarf is heavily favoured as the origin of type Ia supernovae \cite[SNe~Ia;][]{wheeler--81, physics--sne--exp}. There have been many varieties of thermonuclear explosion mechanisms proposed however, and no current scenario is able to reproduce all of the observed properties \cite[see e.g.][]{maoz--14, livio--18}. Additionally, many sub-classes of SNe~Ia have also been identified \cite[see e.g.][]{jha--17, taubenberger--17}. These objects show similarities to SNe~Ia in some ways, but are sufficiently distinct to warrant a separate classification. The peculiarities of many of these objects allow for an exploration of the limits of thermonuclear explosions and the observational signatures they produce.

\par

Perhaps the most extreme of these sub-classes are the `02cx-like' or type Iax supernovae \cite[SNe~Iax; ][]{02cx--orig, 02cx--late--spec, foley--13}. Unlike SNe~Ia, SNe~Iax show great diversity in their light curves and spectra. Photometrically, they are characterised by faint peak absolute magnitudes (up to five magnitudes fainter than SNe Ia) and can show faster declines in their luminosities than SNe Ia \citep{02cx--orig, valenti--09}. Their near-infrared (NIR) light curves also do not show the prominent secondary maximum that is seen in SNe~Ia \citep{02cx--orig}. Previous studies have shown that the secondary maximum results from the recombination of iron group elements (IGEs) from doubly- to singly-ionised as the ejecta expands and cools, and the photosphere recedes into the iron-dominated inner regions \citep{kasen--06a}. It has been suggested that the lack of such a feature in SNe~Iax indicates that iron is present throughout the ejecta and no such sudden change in ionisation state occurs \citep{phillips--07}. Indeed, spectra of SNe~Iax show features due to IGEs are present at all epochs \citep{read--02cx--spectra}. We note however that the absence of a NIR secondary maximum is not exclusive to SNe~Iax (see e.g. \citealt{taubenberger--17}).

\par

The lack of NIR secondary maxima and the presence of features due to IGEs at all epochs has led to the suggestion that the ejecta is well mixed \citep{02cx--late--spec, phillips--07, 05hk--400days}. Such a strongly mixed ejecta is naturally explained in the pure deflagration explosion scenario \citep{reinecke--02a, reinecke--02b}. Here, carbon burning is ignited in the centre of the progenitor white dwarf and propagates sub-sonically throughout the star. Energy released by this thermonuclear burning causes the star to expand. At the same time, the burning front becomes highly turbulent and is accelerated by Rayleigh-Taylor instabilities. It is this turbulence that imprints a large degree of mixing onto the supernova ejecta. The multi-dimensional explosion models of \cite{fink-2014} show a broadly uniform composition to the ejecta (although there are variations along different lines of sight), with burned and unburned material present at all velocities.

\par

Conversely, it has also been suggested that the level of mixing produced by this scenario can not explain the stratification claimed for some of the brightest SNe~Iax. \cite{comp--obs--12z} present observations of SN~2012Z and favour a pulsational delayed detonation \cite[PDD;][]{khokhlov--91b} explosion, which produces a more layered structure to the outer ejecta than that of pure deflagration models \citep{hoeflich-02, dessart--2014b}. In the PDD scenario, the white dwarf experiences expansion during the initial deflagration phase before beginning to contract again. During this contraction, the star may undergo a series of `pulses' that could trigger a detonation and imprint some amount of stratification onto the ejecta. \cite{comp--obs--12z} present several pieces of evidence in favour of the PDD scenario: (i) blue-wings of \ion{Si}{ii}~$\lambda6\,355$ and \ion{Mg}{ii}~$\lambda10\,952$; (ii) a relatively flat velocity evolution for all spectral features; (iii) a `pot-bellied' [\ion{Fe}{ii}] 1.64~$\mu$m profile. In addition, they find that one of the models presented by \citet{hoeflich-02} produces comparable values for the $B$-band peak absolute magnitude, decline rates in the $B$ and $V$ bands, and $B-V$ colour at maximum light when compared to SN~2012Z. With the exception of colour at maximum light however, pure deflagrations also produce comparable values for these light curve properties. Another argument against the ejecta produced by the pure deflagration scenario is presented by \cite{barna--17, barna--18, barna--21}. In these studies, the ejecta composition is mapped for a sample of SNe~Iax by radiative transfer modelling of progressively later epochs. Their analysis shows that the large carbon abundance produced in the inner ejecta layers by pure deflagrations could be difficult to reconcile with observations and the carbon fraction in these regions should be lower. They also state that the mass fraction of other elements decreases towards the outer ejecta. The level of mixing in the ejecta can therefore provide an important constraint on the explosion scenario(s) for the class.

\par

In this study, we present an investigation into the observable consequences of mixing within the ejecta of SNe Iax. We aim to determine what level of mixing (if any) improves the agreement between models and observations of SNe Iax, and whether current observations are able to sufficiently distinguish between mixing and stratification. Our aim is not to argue against or in favour of specific explosion models, but rather to empirically investigate whether a stratified or well-mixed ejecta are capable of reproducing what is observed. We begin in Sect.~\ref{sect:blue-wings} by examining the blue-wing velocities of specific features, which have been claimed as signatures of stratification in the ejecta. In Sect.~\ref{sect:models}, we discuss the models used in this study and their construction. Section~\ref{sect:model_spectra_comp} presents the synthetic spectra of these models and compares them to observations of SN~2012Z and other SNe~Iax. Section~\ref{sect:model_blue_wings} investigates the sensitivity of \ion{Si}{ii}~$\lambda$6\,355 to material in the outer ejecta. The possibility of blending is discussed in Sect.~\ref{sect:blending}. In Sect.~\ref{sect:discussion} we discuss implications for explosion and progenitor scenarios, and finally conclude in Sect.~\ref{sect:conclusions}.

%

\section{Blue-wing velocities}
\label{sect:blue-wings}

\begin{figure}
\centering
\includegraphics[width=\columnwidth]{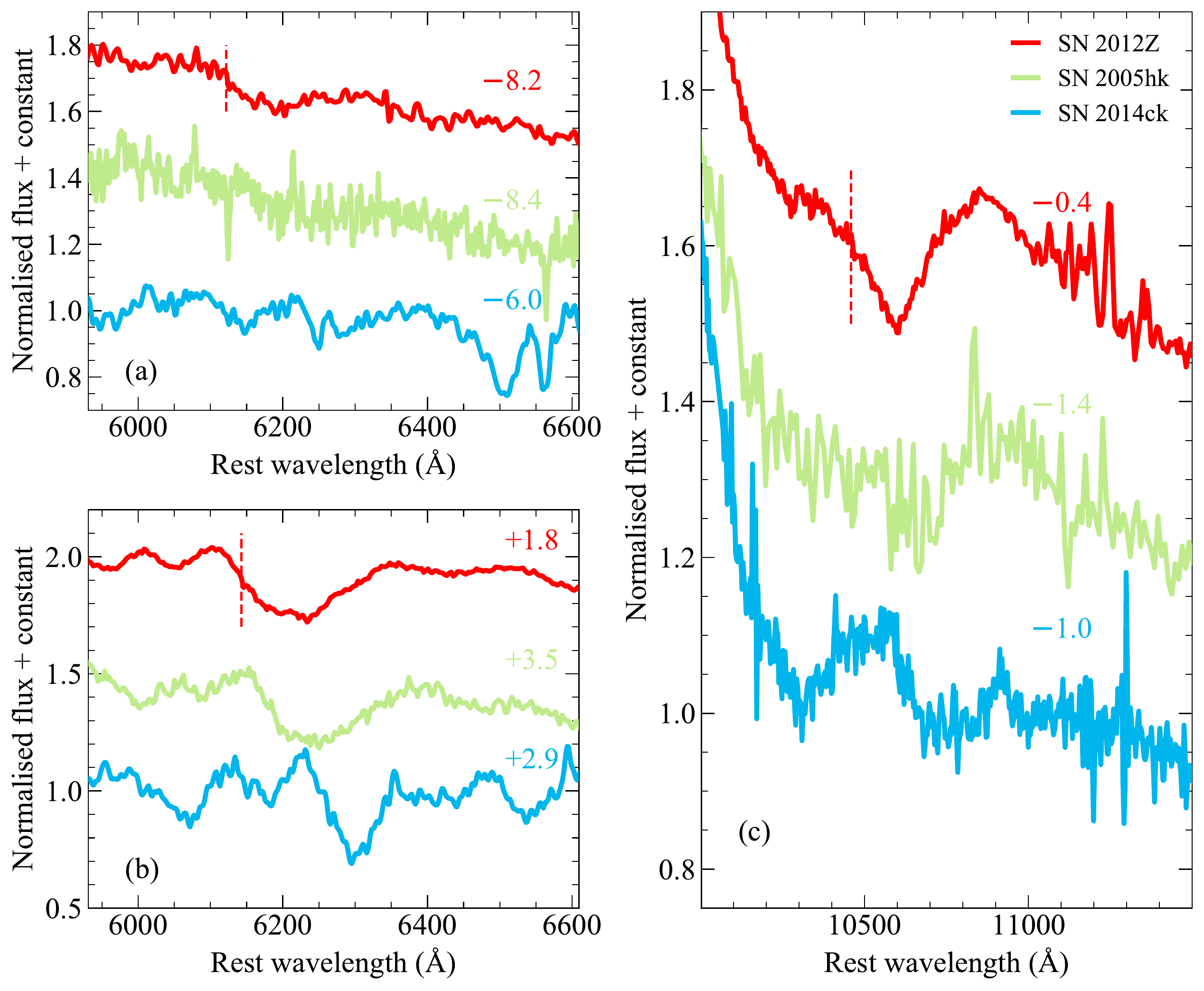}
\caption{Absorption profiles at three epochs for SNe 2005hk, 2012Z, and 2014ck. {\it Panels a} and {\it b} show the \ion{Si}{ii}~$\lambda6\,355$ feature, while {\it Panel c} shows the \ion{Mg}{ii}~$\lambda10\,952$ feature. Dashed lines show the blue-wing velocities determined by \citet{comp--obs--12z} for SN~2012Z from these spectra, assuming identifications of \siii{} and \mgiit{}. Phases for SNe 2005hk and 2012Z are given relative to $B$-band maximum, while SN~2014ck is given relative to $V$-band maximum. }
\label{fig:blue-wing}
\centering
\end{figure}

\begin{figure*}
\centering
\includegraphics[width=\textwidth]{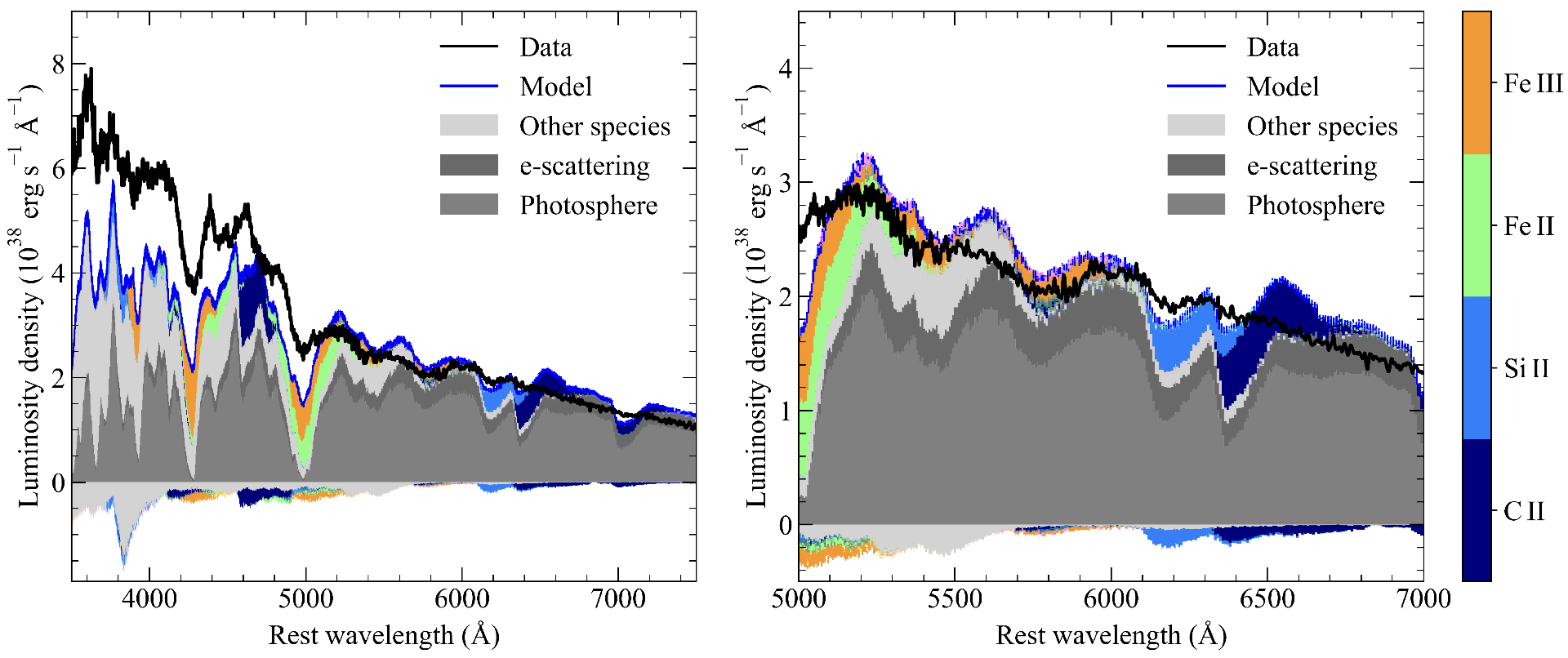}
\caption{Comparison between our best matching \textsc{tardis} model and SN~2012Z approximately eight days before maximum light. {\it Left:} Comparison of the full optical spectra. {\it Right:} Zoom-in of the region surrounding \siii{}. In each panel, we colour code histograms based on the species with which escaping Monte Carlo packets experienced their last interaction. Packets that did not interact are shown as dark grey, while those that experienced only electron-scattering are shown as light grey. Coloured contributions above and below zero show the emitted and absorbed luminosity, respectively, of interacting packets during the simulation for their last interaction. All other species are shown as silver.
}
\label{fig:kromer}
\centering
\end{figure*}

The blue-wing velocity for an absorption feature is determined from the blueshifted edge of the line profile and can provide information on the maximum velocity at which a particular ion is present. \cite{comp--obs--12z} argue that the \ion{Si}{ii}~$\lambda6\,355$ and \ion{Mg}{ii}~$\lambda10\,952$ features in spectra of SN~2012Z show different blue-wing velocities (by $\sim$3\,500~km~s$^{-1}$) and therefore these ions exist in different regions of the ejecta, indicating the ejecta is stratified to some degree. In principle, differences in blue-wing velocities could point to stratification in ionisation state rather than abundance. Two elements could share a similar spatial distribution, but their degree of ionisation differs throughout the ejecta, giving the appearance that the elements are stratified. This point is discussed further in Sect.~\ref{sect:blending}.

\par

Although blue-wing velocities can be powerful tools for interrogating spectra, the large velocities in SNe can make it difficult to extract specific features and determine exactly where an absorption profile ends and the continuum begins. The significant degree of line blending present in SN spectra means that the blue-wing of one absorption profile is often affected by the red-wing of another, or indeed the absorption minimum itself may be affected by the presence of other lines. Therefore, the true maximum velocity of a particular ion may not necessarily be observed and systematic uncertainties may be introduced if the appropriate lines are not identified robustly.

\par

Figure~\ref{fig:blue-wing} demonstrates the challenges present when using absorption profile blue-wings to empirically derive ejecta properties in SNe~Iax. In Fig.~\ref{fig:blue-wing}, we show the observed \siii{} features of SN~2012Z approximately one week before maximum light (Fig.~\ref{fig:blue-wing}(a)) and around maximum (Fig.~\ref{fig:blue-wing}(b)). In Fig.~\ref{fig:blue-wing}(c), we also show the \mgiit{} profile of SN~2012Z around maximum. These three spectra were analysed by \citet{comp--obs--12z} and the measured blue-wing velocities for SN~2012Z from that work are marked as dashed vertical lines. Figure~\ref{fig:blue-wing} shows that the blue-wings of the \ion{Si}{ii}~$\lambda6355$ and \ion{Mg}{ii}~$\lambda10\,952$ absorption profiles are not sharply defined for SN~2012Z. For the $-$8\,d spectrum shown in Fig.~\ref{fig:blue-wing}(a), the location of the blue-wing is highly uncertain as the \siii{} profile gradually rejoins the continuum. The marked velocity does not coincide with any sharp change in the shape of the absorption profile. The associated uncertainty is likely on the order of $\gtrsim$1\,000~\kms{}.  Figure~\ref{fig:blue-wing}(b) also shows how the blue-wing of an absorption feature may become contaminated by the red-wing of another. In this case it would not be possible to accurately determine where absorption from one ion truly ends and another begins. In Fig.~\ref{fig:blue-wing}(b), the blue-wing of the \siii{} profile at maximum light coincides with the red-wing of \ion{Fe}{ii}~$\lambda$6\,247 at $\sim$6\,100~\AA. In other words, there is a turn-over between the two absorption profiles. If this \ion{Fe}{ii} feature were not present, it is conceivable that the \siii{} blue-wing would extend to lower wavelengths. Therefore the presence of another absorption profile could artificially limit the blue-wings to lower velocities. This is also demonstrated for the \mgiit{} feature around maximum light in Fig.~\ref{fig:blue-wing}(c). Here, the absorption is centred around $\sim$10\,600~\AA, however another very broad absorption feature is present around $\sim$10\,200~\AA. In this case, it is not possible to say where the absorption from \mgiit{} truly ends. It is clear however, that there are differences in the purported \siii{} and \mgiit{} profiles around maximum light. In Sect.~\ref{sect:blending}, we discuss how these features may be strongly impacted by blending in the absorption troughs, which further highlights the difficulty in assigning velocities to features in SNe~Iax without detailed modelling of the ejecta or spectral features.

\par

In addition, Fig.~\ref{fig:blue-wing} shows those SNe Iax with optical and NIR spectra at roughly similar phases, SNe 2005hk \citep{phillips--07} and 2014ck \citep{tomasella--2016}. SN~2005hk may have been slightly fainter than SN~2012Z (by $\sim$0.27~$\pm$0.27~mag in the $B$-band; \citealt{comp--obs--12z}) and overall was spectroscopically similar. SN~2014ck was approximately one magnitude fainter than SN~2012Z \citep{tomasella--2016} and showed significantly lower velocities (by $\sim$5\,000~km~s$^{-1}$ around maximum light). It is clear from Fig.~\ref{fig:blue-wing} that both objects show similar blue-wings to SN~2012Z. In all cases, the blue-wing either shows a gradual transition to the continuum or is cut short due to blending with the red-wing of another feature. A robust blue-wing velocity determination is therefore a challenging prospect for these features.

\begin{table*}
\centering
\caption{\textsc{tardis} model parameters and properties}\tabularnewline
\label{tab:model-params}\tabularnewline
\resizebox{\textwidth}{!}{
\begin{tabular}{ccccccccc}\hline
\hline
\multicolumn{4}{c}{Data properties} \vline & \multicolumn{5}{c}{\textsc{tardis} model properties} \tabularnewline
\hline
Data source & Date & MJD & Phase$^{a}$ & Time since       & Luminosity           & Inner boundary         & Blackbody    &    Density \tabularnewline
    &      &     & (days)      & explosion (days) & ($\log$~L$_{\odot}$)  & velocity (km~s$^{-1}$) & temperature (K) & profile \tabularnewline
\hline
\hline
\multicolumn{9}{c}{SN~2005hk} \tabularnewline
\hline
\cite{blondin--12}   & 2005 Nov. 02 & 5\,3676.2 & $-8.4$ & $\phn$6.8 & 8.65 & 9\,600 & 11\,600 & N5def \tabularnewline
\cite{blondin--12}   & 2005 Nov. 06 & 5\,3680.2 & $-4.4$ & 10.8      & 8.85 & 8\,800 & 10\,100 & N5def \tabularnewline
\cite{phillips--07}   & 2005 Nov. 14 & 5\,3688.2 & $+3.6$ & 18.8      & 9.03 & 8\,100 & $\phn$8\,400 & N5def \tabularnewline
\hline
\multicolumn{9}{c}{SN~2012Z} \tabularnewline
\hline
\cite{comp--obs--12z}   & 2012 Feb. 02 & 5\,5959.2 & $-8.2$ & $\phn$6.8 & 8.64 & 10\,000 & 11\,300 & N10def \tabularnewline
\cite{comp--obs--12z}   & 2012 Feb. 12 & 5\,5969.2 & $+1.8$ & 16.8 & 9.04 & $\phn$9\,200 & $\phn$8\,400 & N10def \tabularnewline
\cite{comp--obs--12z}   & 2012 Feb. 16 & 5\,5973.6 & $+5.7$ & 20.7 & 9.07 & $\phn$8\,700 & $\phn$7\,900 & N10def \tabularnewline
\hline
\multicolumn{9}{c}{SN~2014ck} \tabularnewline
\hline
\cite{tomasella--2016}   & 2014 Jul. 01 & 5\,6839.6 & $-6.0$ & 15.0 & 8.62 & $\phn$4\,900 & $\phn$9\,500 & N5def hybrid $\times$10\tabularnewline
\cite{tomasella--2016}   & 2014 Jul. 06 & 5\,6844.5 & $-1.1$ & 19.9 & 8.63 & $\phn$4\,700 & $\phn$8\,300 & N5def hybrid $\times$10 \tabularnewline
\cite{tomasella--2016}   & 2014 Jul. 10 & 5\,6848.5 & $+2.9$ & 23.9 & 8.60 & $\phn$4\,600 & $\phn$7\,500 & N5def hybrid $\times$10 \tabularnewline
\hline
\hline
\multicolumn{9}{l}{$^{a}$ SNe~2005hk and 20212Z relative to $B$-band maximum, SN~2014ck relative to $V$-band maximum.} \tabularnewline
\end{tabular}
}
\end{table*}

\par

To further demonstrate how blue-wings of one feature may be unreliable indicators of the maximum velocity, we use the one-dimensional, Monte Carlo radiative transfer code \textsc{tardis} \citep{tardis, tardis_v2} to model the early spectrum of SN~2012Z eight days before maximum light, which should be the most sensitive spectrum to the outermost ejecta. In Fig.~\ref{fig:kromer}, we compare SN~2012Z and a \textsc{tardis} model using a composition predicted by a pure-deflagration explosion model, which is nearly-uniform (the construction of this model is described in full detail in Sect.~\ref{sect:models}). The input parameters of this model are also given in Table~\ref{tab:model-params}. Figure~\ref{fig:kromer} shows that our model is able to match the velocities and strengths of many of the features observed in SN~2012Z. In particular, the model shows good agreement with the features and continuum from $\sim$5\,000 -- 6\,500~\AA\, (a zoom-in of this region is shown in Fig.~\ref{fig:kromer}).

\par

In Fig.~\ref{fig:kromer} we colour code the model spectrum 
based on the contribution of individual ions to the features. These contributions are calculated through binning escaping Monte Carlo packets by the ion with which they experienced their last interaction. Around $\sim$5,700 -- 6\,100~\AA, many of the escaping Monte Carlo packets had interacted with \ion{Fe}{iii}. Although the model clearly produces a \siii{} feature that is too strong, the overall width is comparable to that of the data. We note that silicon is present throughout the entire model domain. Given this, Fig.~\ref{fig:kromer} shows the difficulty in interpreting blue-wings as the absence of a particular element, compared to a lack of optical depth.

\par

Based on the absorption profiles of the SNe~Iax discussed in this section, we argue that there is sufficient uncertainty in determining robust blue-wing velocities to doubt the claims of stratification in SNe~Iax. We now look to create a series of model spectra for which we can directly investigate the degree of stratification and the impact on the spectra and line profile shapes.

%

\section{Mixing models}
\label{sect:models}

\subsection{Input parameters}
Here we discuss the construction of our model sequence, which is designed to investigate the spectroscopic signatures of mixing within the supernova ejecta. All of our models were calculated using \textsc{tardis} \citep{tardis, tardis_v2}. For each of our \textsc{tardis} simulations, we require a number of input parameters: the density and composition of the ejecta, time since explosion, luminosity, and photospheric velocity.

\begin{figure}
\centering
\includegraphics[width=\columnwidth]{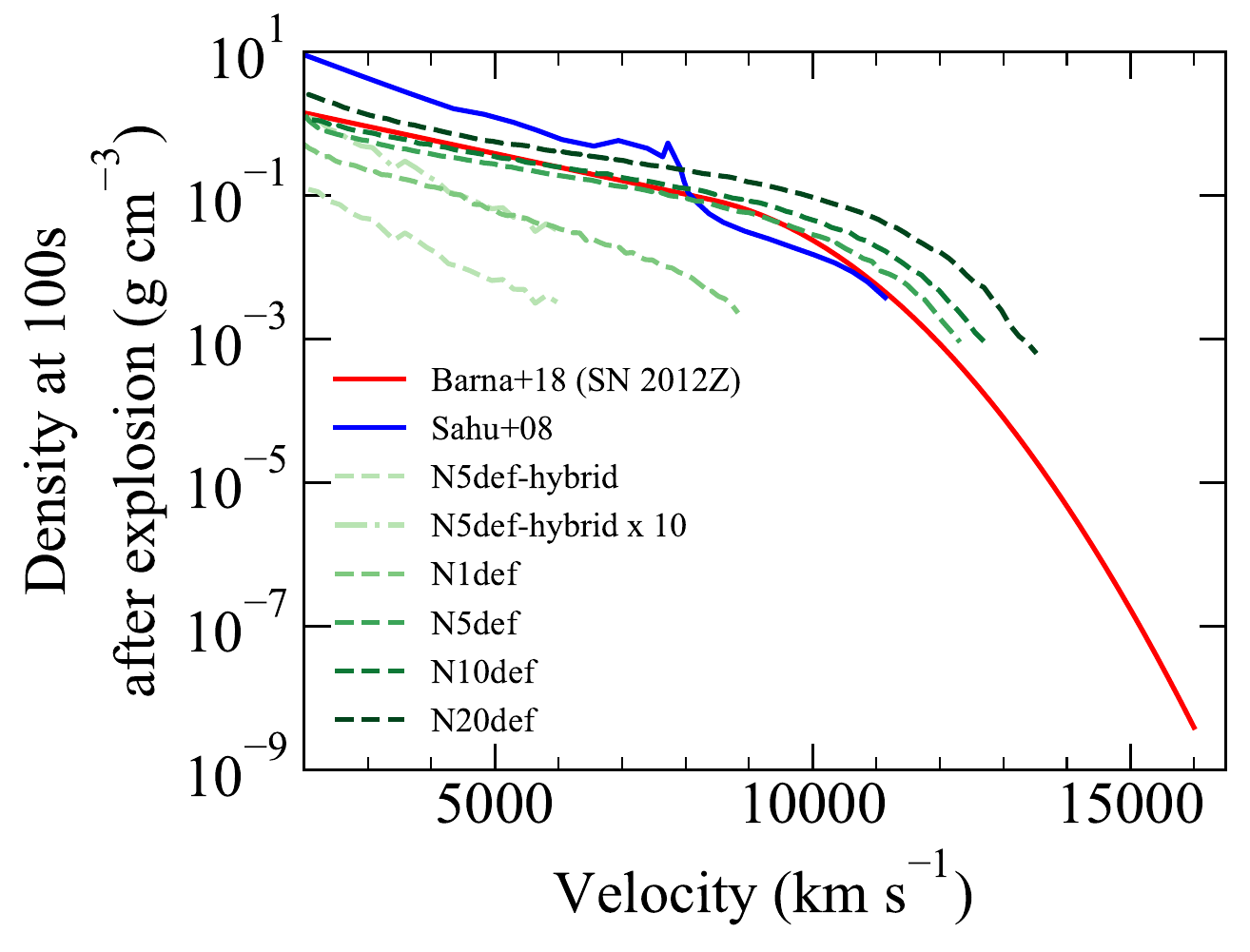}
\caption{Comparison of density profiles used to model SNe~Iax in this work and the literature. \protect\cite{05hk--400days} scale the density profile of the W7 explosion model \protect\citep{nomoto-w7} to a lower kinetic energy to model the spectra and bolometric light curve of SN~2005hk. \protect\cite{barna--18} present density profiles used when modelling a number of SNe~Iax with \protect\textsc{tardis}. The density profile shown here was used to model SN~2012Z. The angle-averaged density profiles of the \protect\cite{fink-2014} and \protect\cite{kromer-15} models used here are also shown, along with the scaled N5def-hybrid density profile used to model SN~2014ck.}
\label{fig:dens}
\centering
\end{figure}

\begin{figure*}
\centering
\includegraphics[width=\textwidth]{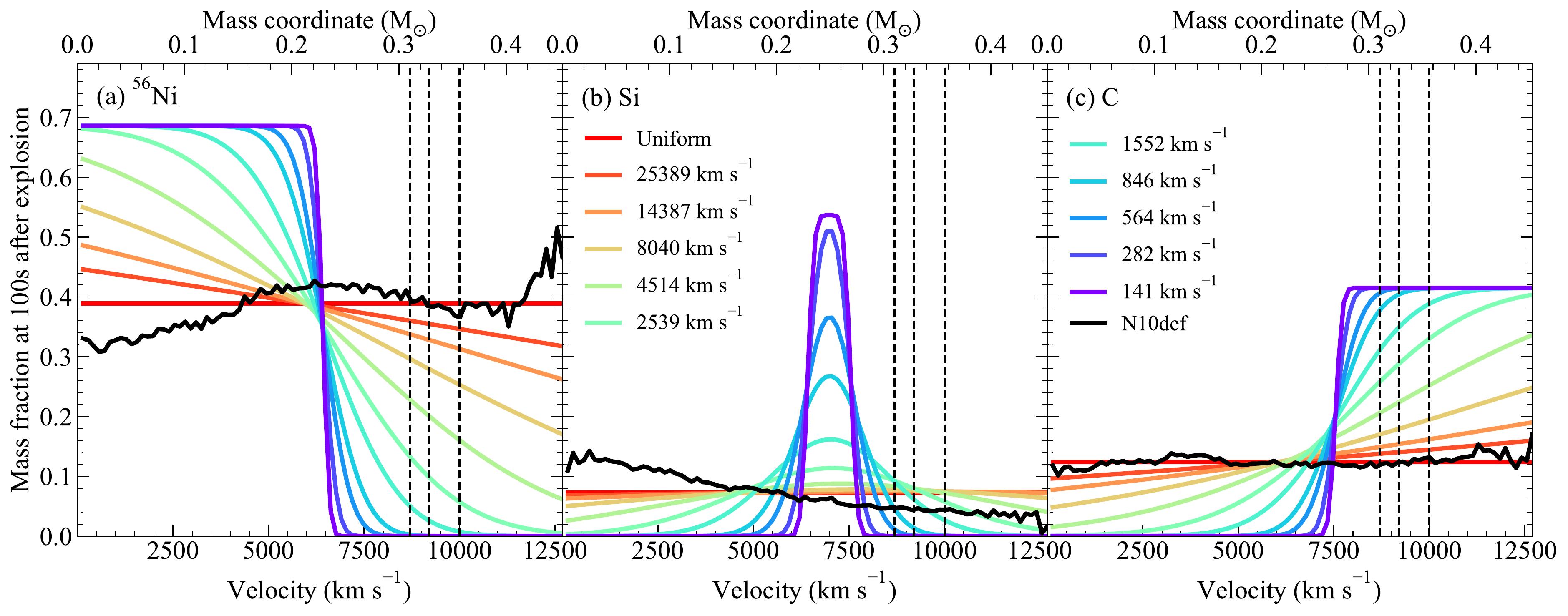}
\caption{Compositions for the toy mixing models used in this work that are based on the N10def explosion model \citep{fink-2014}. Distributions are given for $^{56}$Ni ({\it Panel a}), Si ({\it Panel b}), and C ({\it Panel c}). Mass fractions are shown at 100\,s after explosion. Colours show models with varying levels of mixing, given by the width of the Gaussian smoothing kernel. The angle-averaged composition of the N10def model is shown as a solid black line. Vertical dashed lines show the position of the photosphere for the three epochs presented here. We note that our model spectra at each epoch are not sensitive to material below the photosphere. The entire ejecta structure is shown simply for reference and to demonstrate the construction of the models.
}
\label{fig:model_composition}
\centering
\end{figure*}

\par

A number of different density profiles have been used in the literature to perform similar modelling of SNe~Iax. \cite{05hk--400days} scale the W7 explosion model \citep{nomoto-w7} to a lower kinetic energy (0.3$\times$10$^{51}$~erg cf. 1.3$\times$10$^{51}$~erg) more suitable for SNe~Iax. \cite{barna--18} assume density profiles similar to those predicted by some pure deflagration models \citep{fink-2014}, but add a steeper drop-off in the density at higher velocities. These density profiles are shown in Fig.~\ref{fig:dens} for comparison. 

\par

In this work, we use density profiles predicted from multi-dimensional explosion models presented by \cite{fink-2014} and \cite{kromer-15}. These models have previously been argued to produce synthetic spectra in general agreement with observations of SNe~Iax \citep{kromer-13, kromer-15, 15h, magee--19}. The explosion follows the pure deflagration of either a Chandrasekhar mass carbon-oxygen white dwarf \citep{fink-2014} or a hybrid white dwarf, in which the carbon-oxygen core is surrounded by an oxygen-neon mantle \citep{kromer-15}. In both cases, the strength of the explosion is controlled by the number of sparks (1 -- 1600) used to ignite the deflagration. The number of sparks also gives the name of each model (e.g. N5def was ignited with five sparks). The exact number of sparks used represents a simple method by which the amount of $^{56}$Ni and ejected material can be controlled, although we note that a small number of sparks may represent more physical simulations \citep{nonaka--12}. For models with only a few ignition sparks ($\lesssim$100), the explosion is not sufficient to fully unbind the white dwarf and a remnant is left behind. Such remnants have been predicted to produce narrow P-Cygni features cosistent with late-time spectra of SNe~Iax \citep{02cx--late--spec, 05hk--400days, foley--late--iax}. In addition, \cite{08ha--prog} argue that a remnant may have been directly observed in the case of the extremely low-luminosity SN~Iax, SN~2008ha, but it also cannot be ruled out that the detected source is a surviving companion star. As these deflagration models are multi-dimensional simulations and \textsc{tardis} is a one-dimensional code, we use the densities and compositions provided by HESMA\footnote{\href{https://hesma.h-its.org/doku.php}{https://hesma.h-its.org/doku.php}}\citep{kromer--17}, which have been mapped from the three-dimensional structures of the explosion models into one-dimension by averaging the ejecta density and composition at each radial coordinate in the ejecta. In all of our models, there is no artificial outer boundary added during the \textsc{tardis} simulations. In other words, the maximum velocity in the \textsc{tardis} model comes directly from the explosion simulations. As shown by Fig.~\ref{fig:dens}, for the N5def and N10def models this is $\sim$12\,000~--~12\,500~\kms{}, while the N5def-hybrid model extends to only $\sim$6\,000~\kms{}.

\par

As points of comparison for SNe 2005hk and 2012Z, we select the N5def and N10def models of \citet{fink-2014}, as they produce comparable peak luminosities ($M_{\rm{V}}$ = $-18.24$; $-18.38$) to these SNe~Iax. As previously mentioned, SN~2014ck is an outlier in terms of its peak luminosity and velocities. The N5def-hybrid model of \citet{kromer-15} produces comparable low velocity features to SN~2014ck, however this model is also significantly fainter and faster evolving. For this reason, we arbitrarily scale the density profile by one order of magnitude, which we find produces better agreement with the observations. It remains to be seen how such a SN could be produced within the pure deflagration scenario. The deflagration models presented by \cite{fink-2014} and \cite{kromer-15} have previously been shown to broadly reproduce the light curves and spectra of SNe~2005hk, 2008ha, and 2015H up to shortly after maximum light \citep{kromer-13, kromer-15, 15h}. In all cases however, the models show a faster evolution than is observed. 

\par 

We stress that the purpose of this work is not to provide detailed, individual fits to each object, and indeed not to argue for or against specific explosion models from the literature (such as the N5def model). Instead, this paper is focused on empirical analysis of the data available and the consequences of mixing compared to layering. The question we are trying to address is whether agreement with the observations is improved when one considers a layered ejecta compared to a mixed ejecta. Testing such a question does not preclude the possibility that an existing model cannot perfectly reproduce all of the observations. While some disagreements between observations and the \cite{fink-2014} and \cite{kromer-15} models are apparent, the choice of using density profiles and compositions from these models is a pragmatic decision to reduce the number of free parameters in our model and remove degeneracy that may arise between allowing the density profile and abundances of specific elements to freely vary. 

\par

The remaining free parameters in our \textsc{tardis} simulations (time since explosion, luminosity, and photospheric velocity), were chosen to broadly reproduce the observed spectra of each SN, assuming the angle-averaged density and composition predicted by the explosion model. The input parameters used for our \textsc{tardis} models are given in Table~\ref{tab:model-params}. 

\par

\subsection{Mixing procedure}
We wish to explore the effect of mixing on the synthetic spectral features, but pure deflagration explosion models (such as those upon which our simulations are based) predict an ejecta composition that is well-mixed. This is shown in Fig.~\ref{fig:model_composition} as a black line for the N10def model. Therefore, to investigate the impact of mixing, we treat the distribution of the elements in the ejecta as a free parameter. Specifically, we first create an ejecta structure that is fully stratified. We separate the ejecta into three distinct zones: an inner IGE (Sc -- Ni) zone, a transitional intermediate-mass element (IME; F -- Ca) zone, and an outer carbon-oxygen zone. The mass of each zone is taken directly from the mass produced by each explosion model. For example, the N10def model predicts $\sim$0.25~\mass{} of IGEs is produced during the explosion. Therefore, in our stratified model, we set the inner $\sim$0.25~\mass{}to be composed entirely of IGEs. The relative fractions of elements within these groups (e.g. the fraction of Fe to IGEs) are again taken from the explosion model predictions.

\par

To produce mixed ejecta structures, we use a Gaussian convolution kernel of varying widths. Applying the same method for mixing as \citet{noebauer-17}, the width of the kernel in velocity space is given by 
\begin{equation}
\sigma = k\times w,
\label{eq:gauss}
\end{equation}
where $k$ is the number of cells over which we perform the smoothing and $w$ is the velocity width of each cell. For the N5def, N10def, and N5def-hybrid models, the widths of each cell are $w$ = 138, 141, and 171~km~s$^{-1}$, respectively. Again, these values come directly from the angle-averaged density profiles provided by HESMA. We use ten log-space values for $k$, ranging from one up to twice the number of cells in the model. This range was chosen simply to produce a well sampled space of model compositions from completely stratified to uniform. Figure~\ref{fig:model_composition} shows the distribution of $^{56}$Ni, Si, and C for the N10def models considered here in velocity- and mass-space. These species are representative of the distribution of IGEs, IMEs, and C/O, respectively. Similar profiles are constructed for the other explosion models explored in this work. In addition, we calculate a model for which the ejecta composition is completely uniform. This is shown as a red line in Fig.~\ref{fig:model_composition} for the case of the N10def model.

\par

By changing the composition from stratified to uniform in our models, there will of course also be changes in opacity, radiation field, temperature, and ionisation and excitation state. For the purpose of this work, we are mostly interested in how the physical location of individual ions affects the line profile shape. Therefore, to avoid differences in spectral features being attributable to differences in, for example, ionisation state, we set the temperature profile of all models to be the same as in the model calculated using the unmodified angle-averaged abundances from each explosion model. We have tested this condition by generating models in which we do not impose a temperature profile, but allow this to be calculated by \textsc{tardis} as usual. For our mixed models ($\sigma \gtrsim 1\,000$~km~s$^{-1}$), there are only minor changes in the resultant spectra. For models with a highly-stratified ejecta, as expected, the temperature is generally lower in the outer ejecta (as there are no IGEs in these regions). In general, this leads to weaker features and overall worse agreement compared to the observations, relative to those models in which a temperature profile is fixed.

\par

\begin{figure*}
\centering
\includegraphics[width=\textwidth]{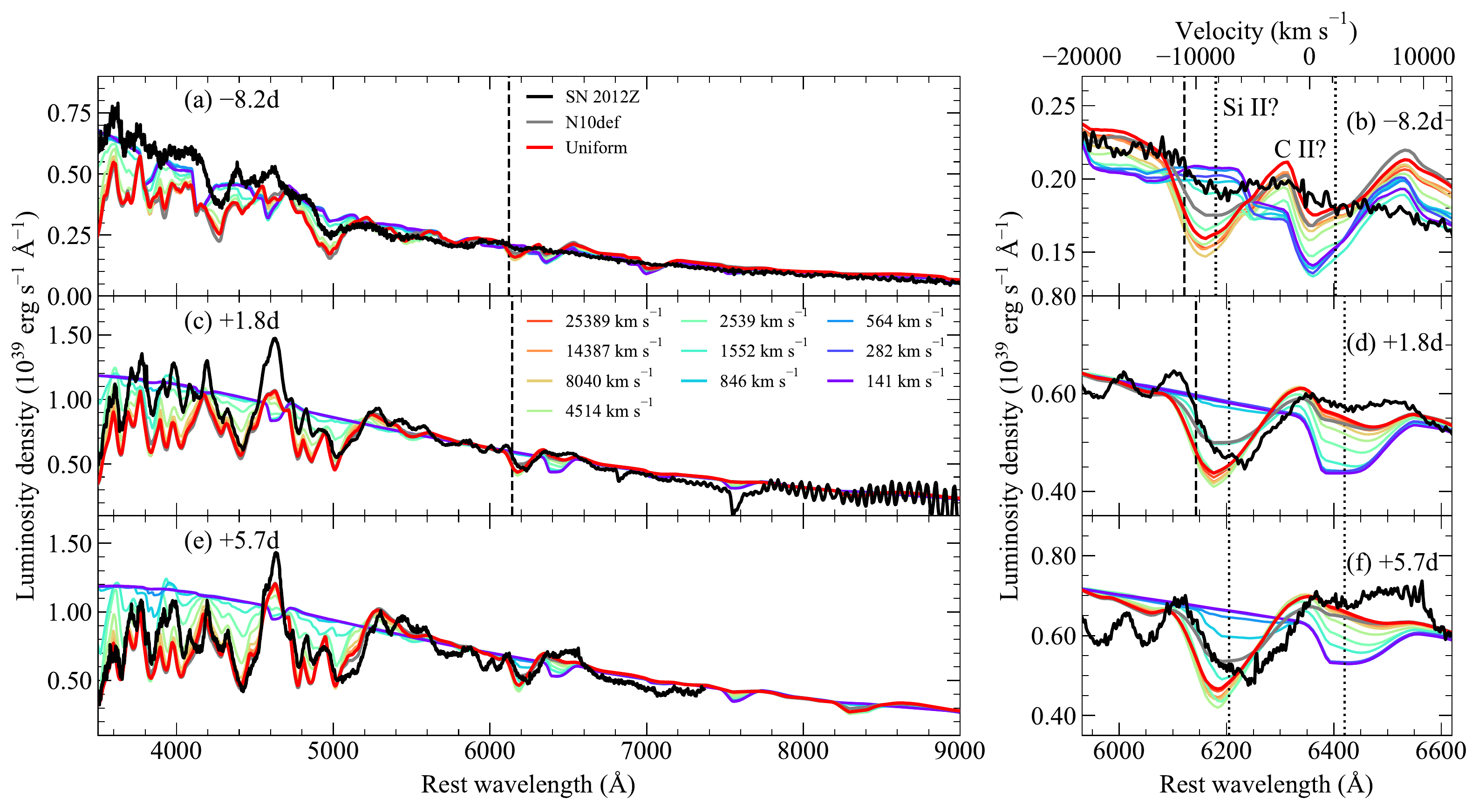}
\caption{Comparison of SN~2012Z to models with varying levels of mixing that are based on the N10def explosion model \citep{fink-2014}. {\it Left:} Comparison of the full optical spectra. {\it Right:} Zoom-in of the region surrounding \siii{}. Velocities give the width of the Gaussian convolution kernel used to construct the mixed ejecta structures, with larger values corresponding to more mixed models. Vertical dashed lines show the position of the \siii{} blue-wings measured by \citet{comp--obs--12z}. Vertical, grey dotted-lines show the approximate locations of potential \ion{Si}{ii} and \ion{C}{II} features during the evolution of SN~2012Z. Velocities in the right panel are given relative to the rest wavelength of \siii{}.
}
\label{fig:12z_comp}
\centering
\end{figure*}


\section{Comparison of mixing models}
\label{sect:model_spectra_comp}

In Fig.~\ref{fig:12z_comp}, we present the spectra calculated for our N10def models with varying levels of mixing. At each epoch, we also show SN~2012Z at a comparable phase. Similar figures for the N5def and N5def-hybrid models compared to SNe~2005hk and 2014ck are shown in the appendix in Figs.~\ref{fig:05hk_comp} and \ref{fig:14ck_comp}, respectively. We first focus our discussion on SN~2012Z in Sect.~\ref{sect:12z} and briefly discuss SNe~2005hk and 2014ck, for which we find similar results, in Sect.~\ref{sect:05hk}. For SN~2012Z, we correct for a total extinction of $E(B-V) = 0.11\pm0.03$~mag \citep{comp--obs--12z} and distance modulus of $\mu = 32.4\pm0.3$~mag \citep{12z--oister}. For SN~2005hk, we assume $E(B-V) = 0.11$~mag and $\mu = 33.46 \pm 0.27$~mag \citep{phillips--07}. Finally, for SN~2014ck we correct for $E(B-V) = 0.48 \pm 0.10$~mag and $\mu = 31.94 \pm 0.15$~mag \citep{tomasella--2016}.

\par

Model comparisons within the literature are often performed using a so-called $\chi$-by-eye metric (e.g. \citealt{hoeflich-02, stehle--05, 11fe--models, smartt--17gfo, watson--19}). While such a procedure is commonly used, the result is that determining the level of agreement between the models and data is an inherently subjective process. Indeed, for parameters that produce only subtle changes in the spectra, it can be difficult to assess what the preferred values are when the models themselves do not perfectly reproduce the data. To help alleviate this issue somewhat, we use SNID \citep{snid} to perform cross-correlations. SNID has been used extensively in the literature to determine the age, redshift, and type of newly discovered SNe and also to compare model spectra to observations (e.g. \citealt{blondin--11, sim--13}). 

\par

During the cross-correlation process, both the SNID input (the observations) and template (the models) are flattened by fitting and dividing through a pseudo-continuum. The spectra are then also normalised such that they have a mean of zero. The result is that SNID is insensitive to the overall luminosity and colour of the spectra, but instead focuses on the relative strengths of spectral features, giving significantly more weight to strong spectral lines compared to weak lines or the continuum shape. The strength of the correlation is determined by both the correlation height-noise ratio, $r$ (which gives the significance of the peak in the cross-correlation function; \citealt{tonry--79}), and amount of wavelength overlap between the input and template spectra, lap. The product of these two values, $r$lap, is used to grade the quality of the fit, with $r$lap $\gtrsim$5 typically considered a `good' match \citep{snid}. 

\par

While some of our models do show $r$lap values $\gtrsim$5, the caveats mentioned previously mean that one cannot simply rely on SNID to determine the overall quality of the comparison. In theory, a systematic offset could be applied by SNID in luminosity and/or velocity space and hence provide a misleading assessment of the level of agreement. Such a velocity offset would indeed be reflected by the redshift determined by SNID, which is part of the fitting procedure. In addition, the model could be significantly brighter or fainter than the data and SNID would not be able to tell due to the removal of the continuum and normalisation. Hence, for our comparisons, we do not rely solely on the $r$lap value or indeed $\chi$-by-eye, as is often done. Rather, a combination of the SNID $r$lap, redshift, and visual inspection provides the best method through which one can more quantitatively asses whether or not one model outperforms another.

\par 

Here, we present $r$lap values calculated across the entire observed spectrum and a more restricted wavelength range centred around the \siii{} profile ($5\,800$~\AA~$\textless~\lambda~\textless~6\,800$~\AA). These are given in Tables~\ref{tab:snid_12z}~\&~\ref{tab:snid_12z_si} for SN~2012Z and our N10def models. We note that for the correlations performed centred around the \siii{} profile, the strength of the correlation ($r$) may be relatively high, but the limited wavelength coverage (lap $\sim$ 0.15) leads to small values of $r$lap, which are typically graded by considering much larger wavelength ranges ($\Delta\lambda\gtrsim2\,500$~\AA; \citealt{snid}). We again note that within SNID, a systematic velocity offset may be applied in the form of the redshift. By comparing against the measured redshifts of SNe however, one can easily judge whether such a systematic offset is required. Therefore, in Tables~\ref{tab:snid_12z}~\&~\ref{tab:snid_12z_si} we provide the redshift determined for SN~2012Z (along with the associated uncertainty), which may also be used to aid in qualitatively assessing the level of agreement when compared to the measured redshift of $z = 0.007$ \citep{comp--obs--12z, 12z--oister}.

\par

\subsection{SN~2012Z}
\label{sect:12z}

In Fig.~\ref{fig:12z_comp}, we show comparisons between SN~2012Z and our mixing models. While it is clear from Fig.~\ref{fig:12z_comp} that the model spectra do broadly reproduce the data, there are noticeable differences and some features are not well reproduced by any of the models. The similarities and discrepancies are discussed in more detail below. The overall level of agreement however, is comparable to that for models that have been previously used to argue in favour of pure deflagrations \citep{kromer-13}, pulsational delayed detonation models \citep{comp--obs--12z}, or indeed a layered ejecta \citep{barna--18}. To make a definitive statement regarding the nature of the ejecta would require perfect agreement with the observations, which is an impossible task due to myriad uncertainties during the modelling process, including the initial conditions of the white dwarf and explosion physics. Therefore the model and observed spectra demonstrate the complexity of the problem in determining whether the ejecta must be mixed or layered. In what follows, we focus on discussing the impact of mixing compared to layering and whether this is sufficient to improve agreement with the observations. Of course, such an exercise does not imply the correct model, as is the case for all similar comparisons, but demonstrates how sensitive the observations may be to, for example, mixing.

\par

Figure~\ref{fig:12z_comp}(a) shows our N10def models, with different levels of mixing, at seven days after explosion compared to SN~2012Z at a roughly equivalent epoch, eight days before $B$-band maximum. It is clear from Fig.~\ref{fig:12z_comp}(a) that models with little or no mixing ($\sigma \lesssim 1\,500$~\kms{}) do not reproduce many of the features in the observed spectrum. At seven days after explosion, these models contain only C/O above the photosphere (see Fig.~\ref{fig:model_composition}) and therefore do not produce the \ion{Si}{ii}~$\lambda$6\,355 feature or indeed the strong features due to IGEs visible at shorter wavelengths ($\lesssim$5\,000~\AA). The lack of such features, and indeed the production of generally only weak lines, is also demonstrated by the low $r$lap values, compared to our other models, and the need for a high redshift, as shown in Tables~\ref{tab:snid_12z}~\&~\ref{tab:snid_12z_si} .

\par

\begin{table}
\centering
\caption{SNID $r$lap correlation coefficients for SN~2012Z.}\tabularnewline
\label{tab:snid_12z}\tabularnewline
\resizebox{\columnwidth}{!}{
\begin{tabular}{cccccccc}\hline
\hline
Model & \multicolumn{2}{c}{2012 Feb. 02} & \multicolumn{2}{c}{2012 Feb. 12} & \multicolumn{2}{c}{2012 Feb. 16} & Mean \tabularnewline
(km~s$^{-1}$) &  & & & & &  & $r$lap \tabularnewline
\hline
\hline
& $r$lap & $z$ & $r$lap & $z$ & $r$lap & $z$ & \tabularnewline
\hline
$\phn\phn$141 & 1.33 & 0.022(0.030) & 1.17 & -0.030(0.020) & 1.15 & -0.027(0.022) & 1.22 \tabularnewline
$\phn\phn$282 & 1.30 & 0.022(0.031) & 1.55 & -0.030(0.018) & 1.32 & -0.027(0.020) & 1.39 \tabularnewline
$\phn\phn$564 & 1.27 & 0.021(0.035) & 1.22 & -0.029(0.019) & 1.20 & -0.028(0.023) & 1.23 \tabularnewline
$\phn\phn$846 & 1.14 & 0.024(0.032) & 1.33 & -0.031(0.019) & 0.70 & -0.045(0.038) & 1.06 \tabularnewline
$\phn$1\,552  & 1.35 & 0.017(0.030) & 2.11 & -0.003(0.019) & 0.74 &  0.008(0.022) & 1.40 \tabularnewline
$\phn$2\,539  & 3.43 & 0.010(0.016) & 5.05 & 0.001(0.008)  & 2.16 &  0.009(0.016) & 3.55 \tabularnewline
$\phn$4\,514  & 5.69 & 0.010(0.010) & 6.88 & 0.004(0.006)  & 5.01 &  0.010(0.008) & 5.86 \tabularnewline
$\phn$8\,040  & 6.62 & 0.010(0.008) & 7.19 & 0.004(0.005)  & 7.11 &  0.010(0.006) & 6.97 \tabularnewline
14\,387       & 6.90 & 0.010(0.008) & 7.12 & 0.005(0.006)  & 7.93 &  0.010(0.006) & 7.32 \tabularnewline
25\,389       & 7.68 & 0.010(0.008) & 7.12 & 0.005(0.006)  & 8.09 &  0.010(0.006) & 7.63 \tabularnewline
Uniform       & 7.13 & 0.010(0.008) & 7.12 & 0.005(0.006)  & 8.16 &  0.010(0.005) & 7.47 \tabularnewline
N10def        & 7.93 & 0.009(0.008) & 6.82 & 0.005(0.007)  & 7.87 &  0.009(0.006) & 7.54 \tabularnewline
\hline
\hline
\end{tabular}
}
\end{table}

\begin{table}
\centering
\caption{SNID $r$lap correlation coefficients for SN~2012Z surrounding the \siii{} profile, where $5\,800$~\AA~$\textless~\lambda~\textless~6\,800$~\AA.}\tabularnewline
\label{tab:snid_12z_si}\tabularnewline
\resizebox{\columnwidth}{!}{
\begin{tabular}{cccccccc}\hline
\hline
Model & \multicolumn{2}{c}{2012 Feb. 02} & \multicolumn{2}{c}{2012 Feb. 12} & \multicolumn{2}{c}{2012 Feb. 16} & Mean \tabularnewline
(km~s$^{-1}$) &  & & & & &  & $r$lap \tabularnewline
\hline
\hline
& $r$lap & $z$ & $r$lap & $z$ & $r$lap & $z$ & \tabularnewline
\hline
$\phn\phn$141 & 0.23 & -0.009(0.000) & 0.06 &  0.073(0.065) & 0.09 & 0.065(0.074) & 0.13 \tabularnewline
$\phn\phn$282 & 0.26 & -0.009(0.000) & 0.06 &  0.071(0.068) & 0.07 & 0.069(0.086) & 0.13 \tabularnewline
$\phn\phn$564 & 0.26 & -0.009(0.000) & 0.05 &  0.075(0.067) & 0.07 & 0.078(0.063) & 0.13 \tabularnewline
$\phn\phn$846 & 0.14 & -0.009(0.000) & 0.06 &  0.075(0.061) & 0.13 & 0.009(0.016) & 0.11 \tabularnewline
$\phn$1\,552  & 0.12 & 0.011(0.014)  & 0.58 &  0.011(0.018) & 1.15 & 0.012(0.017) & 0.62 \tabularnewline
$\phn$2\,539  & 0.87 & 0.011(0.013)  & 1.69 &  0.012(0.013) & 1.99 & 0.014(0.013) & 1.52 \tabularnewline
$\phn$4\,514  & 1.25 & 0.012(0.012)  & 3.07 &  0.013(0.009) & 3.76 & 0.014(0.009) & 2.69 \tabularnewline
$\phn$8\,040  & 1.31 & 0.012(0.011)  & 3.63 &  0.013(0.008) & 4.82 & 0.014(0.007) & 3.25 \tabularnewline
14\,387       & 1.41 & 0.011(0.011)  & 4.32 &  0.013(0.007) & 5.40 & 0.014(0.007) & 3.71 \tabularnewline
25\,389       & 1.45 & 0.011(0.011)  & 3.92 &  0.013(0.008) & 5.37 & 0.014(0.007) & 3.58 \tabularnewline
Uniform       & 1.42 & 0.011(0.011)  & 4.09 &  0.013(0.008) & 5.51 & 0.014(0.007) & 3.67 \tabularnewline
N10def        & 1.20 & 0.010(0.012)  & 3.07 &  0.011(0.010) & 4.19 & 0.012(0.009) & 2.82 \tabularnewline
\hline
\hline
\end{tabular}
}
\end{table}

Models that are well mixed ($\sigma \gtrsim 4\,000$~\kms{}) provide improved agreement with SN~2012Z as these models do contain some amount of burned material (IMEs and IGEs) above the photosphere, in addition to unburned C/O. The $r$lap values for these models are all above 5. As previously mentioned, this would be considered a `good' spectroscopic match when classifying supernovae \citep{snid}. The redshift determined by SNID is also comparable to the measured value for SN~2012Z, thereby indicating there is minimal velocity offset between the models and observations. For $\lambda\lesssim$5\,000~\AA, the mixed models in general match the shapes and strengths of the observed features, although the overall flux level is somewhat lower in the models (as previously mentioned, SNID removes the continuum flux and only considers spectral features). We note that similar disagreement was also the case for the N5def model spectra presented by \citet{kromer-13}. Some of the discrepancy between the mixed model spectral features and the observations can be attributed to the strong C/O features predicted. In particular, the models show prominent \ion{C}{ii}~$\lambda$4\,270, $\lambda$4\,740, and $\lambda$6\,580 features. \citet{12bwh} have previously shown how a reduced carbon abundance relative to the pure deflagration models of \citet{fink-2014} provides improved agreement with the SNe~Iax SN~2005hk and PS1-12bwh. Reducing the carbon abundance in our mixed models would likely also improve spectroscopic agreement with these SNe~Iax. See Sect.~\ref{sect:carbon} for further discussion of carbon in SNe~Iax.

\par

In Fig.~\ref{fig:12z_comp}(b), we show a zoom in of the \ion{Si}{ii}~$\lambda$6\,355 feature at eight days before maximum light. Figure~\ref{fig:12z_comp}(b) shows that the \siii{} velocities in most model spectra are slightly higher than observed for SN~2012Z. Although again the best-fitting redshifts determined by SNID (see Table~\ref{tab:snid_12z_si}) are comparable to that measured for SN~2012Z and consistent within the uncertainty. The shapes of the \ion{Si}{ii}~$\lambda$6\,355 features are generally similar to SN~2012Z for our mixed models. This is again demonstrated by the $r$lap values given in Table~\ref{tab:snid_12z_si}. Here, we note that the small wavelength region being fitted (lap $\textless$ 0.15) produces similarly small $r$lap values, however the strength of the correlation is high for the mixed models ($\sigma\gtrsim2\,500$~\kms{}). While the shape of the \siii{} feature generally agrees with the mixed models, it is weaker in the data. The \ion{C}{ii}~$\lambda$6\,580 feature is also clearly weaker in the data compared to the models. Again, decreasing the carbon abundance relative to the \cite{fink-2014} deflagration models may improve agreement. Alternatively, given that the \siii{} feature is also too strong, this could point to the density being too high in this region of the ejecta.

\par

The level of agreement between our models and SN~2012Z would indicate that stratification of silicon in the very outermost ejecta is not a requirement to match the observations at these epochs. Instead, the spectrum at this epoch (approximately one week before maximum) may simply be insensitive to the silicon present at high velocities in our model. This point is explored further in Sect.~\ref{sect:model_blue_wings}. The strength of the \siii{} feature in our angle-averaged N10def model is lower than for the mixed models simply due to the reduced silicon abundance above the photosphere (see Fig.~\ref{fig:model_composition}), although the distribution is also approximately uniform.

\par

Interestingly, we find that at approximately one week after explosion our well-mixed ($\sigma \gtrsim 4\,000$~\kms{}) and uniform N10def models (and indeed the N5def models, and N5def-hybrid models with $\sigma \gtrsim 3\,000$~\kms{}) all produce spectra that are very similar to each other, and to models with the angle-averaged abundances. There is little change between the shapes of line profiles apart from the different overall strengths of the features, which correlates with the mass fraction above the photosphere. Similarly there are only small changes in the $r$lap values determined by SNID. This is in spite of the fact that for the angle-averaged N10def models the $^{56}$Ni mass fraction is increasing in the very outermost regions of the ejecta, while for all other mixing models it is either decreasing or constant (see Fig.~\ref{fig:model_composition}). This demonstrates that even at eight days before maximum light, it is not possible to properly discriminate between these cases. The primary differences between the model spectra are due to the amount of burned material above the photosphere, but models at this epoch are unable to constrain the distribution of this material further. This is in contradiction to the models presented by \citet{barna--18} for SN~2012Z (and SN~2005hk), which are argued to require a decreasing fraction of IGE towards the outer ejecta. Our mixing models show that earlier spectra (more than one week before $B$-band maximum) are required to determine whether such a gradient in composition is truly required. We also note that our models are likely to be more sensitive to the material in the outer ejecta compared to \cite{barna--18}, due to the higher density above the photosphere (see Fig.~\ref{fig:dens}).

\par

At 16.8\,d after explosion (Fig.~\ref{fig:12z_comp}(c)), our models with little or no mixing ($\sigma \lesssim 1\,500$~\kms{}) do not match the features of SN~2012Z around maximum light. These models contain very little burned material above the photosphere and generally show a continuum with few features. For these models, the $r$lap values are again very low ($\lesssim$1.6). As was the case for our 6.8\,d spectrum, the mixed models ($\sigma \gtrsim 4\,000$~\kms{}) agree more closely with the relative strengths and shapes of features at shorter wavelengths in SN~2012Z and show higher $r$lap values. At maximum light, SN~2012Z has now started to develop weak absorption features due to \ion{Fe}{ii}~$\lambda$6\,149 \& $\lambda$~6\,247 that are clearly apparent towards the blue end of the \siii{} profile (see Fig.~\ref{fig:12z_comp}(d)). These features are clearly significantly weaker in our \textsc{tardis} models, but still slightly visible. Indeed, these features are most prominent in our angle-averaged N10def model, for which the fraction of iron to silicon above the photosphere is larger than in our mixed models. The \siii{} profile in our mixed models show good agreement with SN~2012Z, as indicated by the relatively high $r$lap values (again we note this is affected by the small wavelength region, lap$\textless$0.15). This would further indicate that it is not a requirement that silicon is stratified as the N10def model and our mixed models contain silicon throughout the ejecta. Our well-mixed, uniform, and angle-averaged N10def models also show better agreement with the weak \ion{C}{ii}~$\lambda$6\,580 feature at this epoch.

\par

The \siii{} velocities in the mixed models are however slightly higher than the absorption profile centred around $\sim$6\,200~\kms{} in SN~2012Z at $+1.8$\,d after maximum. This is reflected in the best-fitting redshift determined by SNID, which has increased slightly relative to the first epoch but remains comparable to SN~2012Z within the uncertainties (see Table~\ref{tab:snid_12z_si}). Alternatively, rather than the model \siii{} profile being formed at velocities that are too high, another possibility is that the observed \siii{} profile is blended with another feature. From analysing the escaping packets, our \textsc{tardis} models indicate some small amount of \ion{Fe}{ii} absorption around this wavelength region and blending of \siii{} with \ion{Fe}{ii} around maximum light has also previously been suggested by \cite{obs--2011ay} for the bright SN~Iax, SN~2011ay. If the model is simply missing significant \ion{Fe}{ii} absorption in this region to match the observations, this would be consistent with the model also producing \ion{Fe}{ii}~$\lambda$6\,149 \& $\lambda$~6\,247 features that are too weak. This point is discussed further in Sect.~\ref{sect:blending}.

\par

In Fig.~\ref{fig:12z_comp}(e), we show our models at approximately three weeks after explosion compared to SN~2012Z at an approximately equivalent epoch -- six days after maximum light. Figure~\ref{fig:12z_comp}(e) and Table~\ref{tab:snid_12z} demonstrate that our mixed models still produce spectra in closer agreement with SN~2012Z over models with little or no mixing. For our most highly mixed models, the $r$lap values given by SNID indicate a good match ($\gtrsim$5.4) even when considering the limited wavelength range (Table~\ref{tab:snid_12z_si}), while the redshifts and luminosity are also in agreement with the observations. The angle-average N10def model and some less mixed models also show reasonably high cross-correlation significance.

\par

\subsection{SNe~2005hk and 2014ck}
\label{sect:05hk}

Both SNe~2005hk and 2014ck show similar trends when compared to our mixed models as SN~2012Z, therefore we briefly consider these objects here. Pre-maximum spectra of SN~2005hk are shown in Figs.~\ref{fig:05hk_comp}(a) \& (c), while SN~2014ck is shown in Figs.~\ref{fig:14ck_comp}(a)  \& (c). As was the case for SN~2012Z, those models with little or no mixing ($\sigma \lesssim 1\,500$~\kms{}) do not contain any burned material above the photosphere and hence are unable to reproduce the observed spectra. Models that are well mixed show improved agreement and $r$lap values that would indicate a `good' match when considering the full spectra. Again, we find that the \ion{C}{ii} features are generally somewhat too strong. Considering only the region surrounding \siii{}, the $r$lap values also indicate that the well-mixed models perform better than the stratified models.

\par

The post-maximum spectrum of SN~2005hk is shown in Fig.~\ref{fig:05hk_comp}(e) and that of SN~2014ck is shown in Fig.~\ref{fig:14ck_comp}(e). Again, models with little or no mixing do not contain burned material above the photosphere and generally show a continuum with a few features.  As with the earlier epochs, we again find that well-mixed models produce the highest $r$lap values when considering either the full spectrum or the region surrounding \siii{}.

\subsection{Summary of mixing comparisons}

Comparing our models with different levels of mixing to observations of SNe~Iax ranging from approximately one week before maximum light to one week after, we find that models that are heavily mixed (i.e. contain IGEs and IMEs throughout the ejecta) provide good agreement with the relative strengths and shapes of spectral features. While some disagreements compared to the data remain, likely due to the composition and density profile used, it is clear that a well-mixed ejecta provides significantly better agreement than a stratified one. In combination with the typical visual inspection, we use SNID to aid in quantification of the agreement through the $r$lap parameter and fitted redshift. In all cases we find that our heavily mixed models produce larger correlation coefficients, relative to the stratified models, when considering both the entire optical spectrum and only the wavelengths surrounding the \siii{} profile. In addition, models calculated using the angle-averaged compositions of pure deflagration explosion models (which also contain a near-uniform silicon composition) show similarly good agreement with observations and in particular the shape of the \siii{} blue-wings. This would indicate that a stratified silicon abundance, as suggested by \citet{comp--obs--12z}, is not a requirement to match the observations. Although there are disagreements, in particular a few features are too strong in our earliest spectrum, it is clear that many of the disagreements are reduced as the ejecta is progressively mixed. Given the level of agreement between our mixed models, the angle-averaged N10def model, and SN~2012Z, we argue that there is insufficient evidence to suggest that the ejecta of SN~2012Z must be layered or is inconsistent with a well-mixed ejecta, such as that predicted by pure deflagration explosions.

%

\section{Stratification of silicon versus steep density profiles}
\label{sect:model_blue_wings}

\begin{figure}
\centering
\includegraphics[width=\columnwidth]{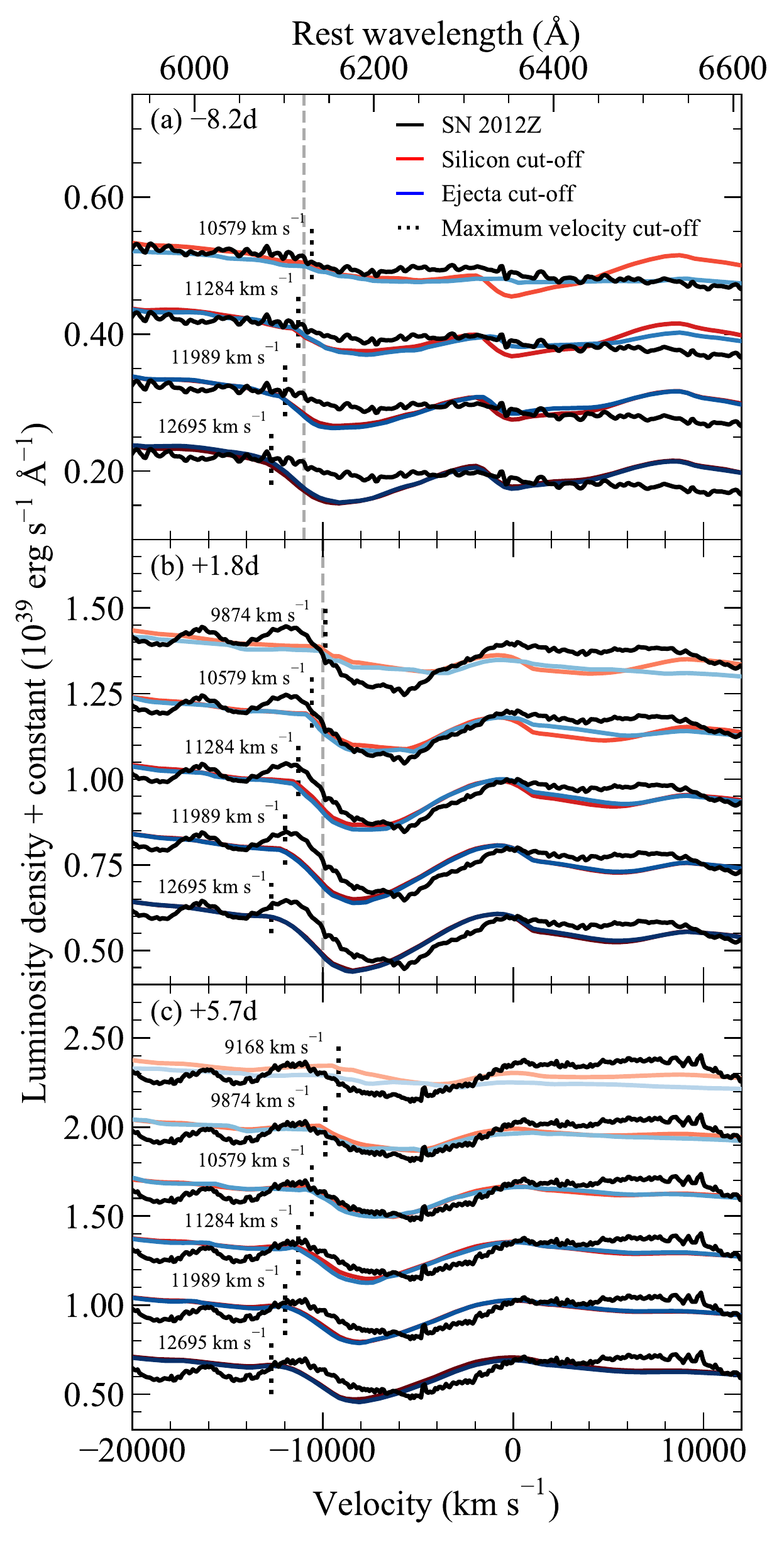}
\caption{Comparison of SN~2012Z to models in which the silicon abundance or entire ejecta is removed for the outer ejecta. Spectra are shown at $-$8\,d ({\it Panel a}), $+$2\,d ({\it Panel b}), and $+$6\,d ({\it Panel c}) relative to $B$-band maximum. The velocity above which these cut-offs are applied are given by vertical dotted lines. Models for which the outer silicon is removed are shown in shades of red, while models for which the outer ejecta is removed are shown in shades of blue. Vertical dashed lines (grey) show the blue-wing velocities of \siii{} measured by \citet{comp--obs--12z}. Vertical offsets have been applied for clarity, however the same offset is applied to all spectra within each comparison.}
\label{fig:12z_comp_si}
\centering
\end{figure}

Our models show that a well-mixed, or even uniform, composition for the ejecta provides good agreement with the relative strengths and shapes of many of the features observed in SN~2012Z. Here, we perform a more direct analysis on the shape of the \siii{} profile by investigating whether stratifying solely the silicon abundance produces improved agreement.

\par

We first take the composition of the angle-averaged N10def model, which we have shown to reproduce the spectrum of SN~2012Z reasonably well (see Sect.~\ref{sect:model_spectra_comp}). Beginning at the outer ejecta and moving progressively inwards, we construct a new, separate series of models by setting the silicon abundance equal to zero in each model cell. The removed silicon fraction is replaced with oxygen, which acts as a filler. This process ensures that the relative fractions of all other elements except silicon and oxygen remain the same. We also construct a set of models for which we add an outer boundary into our \textsc{tardis} simulations and move it progressively inwards. In other words, the ejecta above a certain velocity is neglected from the simulation. Together, these two sets of models allow us to investigate specifically whether agreement is improved by stratification of silicon or if the models are simply insensitive to this high velocity material at certain epochs. In Fig.~\ref{fig:12z_comp_si}, we show a comparison between the \siii{} profiles of SN~2012Z and our models with either a silicon or ejecta cut-off at various velocities. The vertical black dotted lines show the location of the cut-off applied for each model. 

\par

In Fig.~\ref{fig:12z_comp_si}(a), we show the spectrum of SN~2012Z eight days before maximum light. It is clear that, for this epoch, removing either the silicon abundance or ejecta above $\sim$12\,000~\kms{} produces little change in the spectrum. Therefore, even at this early phase, the model spectrum is not sensitive to material above this velocity.  We note that this also shows how blue-wing velocities are unreliable measures of the ejecta composition and come with relatively large uncertainties. For silicon cut-offs of $\gtrsim$12\,000~\kms{} there is virtually no change in the shape of the blue-wing. Silicon could be present at higher velocities in the SN ejecta (in this case silicon extends up to the maximum velocity in the model, $\sim$12\,700~\kms{}), but the optical depth is simply not high enough to impact the spectrum. Therefore it would not be possible to discriminate between stratification or mixing in this case. Applying a silicon or ejecta cut-off of $\sim$11\,300~\kms{}, there is a more dramatic change in the model spectrum. Both models show similar and weaker \siii{} profiles, in better agreement with the observed spectrum. The ejecta cut-off model however, also shows weaker \ion{C}{ii}~$\lambda$6\,580 absorption, again in better agreement with the observations. Applying an even lower velocity ejecta cut of $\sim$10\,600~\kms{} produces significantly weaker features that are still comparable to the data. In this case, the model includes only $\sim$600~\kms{} above the photosphere. The changes in the strengths of these features are not due to changes in ionisation state as the photospheric temperatures in all models are approximately equal.

\par

Figure~\ref{fig:12z_comp_si}(b) shows the spectrum of SN~2012Z shortly after maximum light. Again we find that cutting off the silicon abundance or ejecta above $\sim$12\,000~\kms{} does not have a significant impact on the spectrum. Figure~\ref{fig:12z_comp_si}(b) also shows how for some models, it is not possible to determine a maximum silicon velocity, due to the presence of the \ion{Fe}{ii}~$\lambda$6\,247 feature -- silicon is present at high velocities, but this overlaps with the red-wing of \ion{Fe}{ii}~$\lambda$6\,247. Compared to the observed spectrum, a cut off of $\sim$10\,000~\kms{} removes too much material and produces significantly weaker \siii{} absorption. For cut-offs around $\sim$10\,600~\kms{}, we find improved agreement. Removing the high velocity material results in an apparent shift of the absorption feature around $\sim$6\,200~\AA\, to lower velocities. Around maximum light, this feature could be due to a blend of \ion{Si}{ii} and an additional species (see Sect.~\ref{sect:blending}). Therefore by removing the high velocity silicon, the relative strength of the other absorption feature increases and the two absorption troughs are more easily distinguished, which is consistent with the weak double-trough feature observed in SN~2012Z. Both the silicon and ejecta cut-off models show similar levels of agreement with the observed \siii{} profile, but the ejecta cut-off model also shows weaker \ion{C}{ii}~$\lambda$6\,580 absorption. Moving later to the spectrum approximately one week after maximum light, we again find that the best agreement is for models with an ejecta cut-off between $\sim$10\,600 -- 11\,300~\kms{}.

\par

Our models with various silicon or ejecta cut-offs applied show that it is not the case that a stratified silicon distribution is a requirement to match SN~2012Z. Rather, models for which the high velocity ejecta has been removed provide better overall agreement in terms of the strengths, shapes, and velocities of the \ion{Si}{ii} and \ion{C}{ii} features. This demonstrates that a steeper density profile relative to the N10def pure deflagration model, in which the high velocity ejecta contributes less to the observed spectrum, may be the preferred ejecta configuration. We are unable to place constraints on the necessary composition for such low density and high velocity material. Earlier observations (than approximately eight days before maximum light) would be required to determine whether this high velocity material must be stratified or is also consistent with a well-mixed composition.

%

\section{Blended absorption features}
\label{sect:blending}

In this work, we have demonstrated that reported blue-wings cannot be easily interpreted as indicating layering. Furthermore, our models suggest that the data at these epochs are generally insensitive to modest amounts of layering and are also consistent with a well-mixed ejecta. Neither scenario could therefore be robustly excluded on the basis of the data currently available. Nevertheless, it is clear that there are significant differences between the shapes of the observed \siii{} and \mgiit{} profiles around maximum light (see Fig.~\ref{fig:blue-wing}). Indeed, the models presented thus far have focused solely on optical spectra, however \cite{comp--obs--12z} use a combination of optical and NIR spectra to suggest that the ejecta of SN~2012Z is layered. While it is plausible that such differences may be due to layering, it has also been suggested that, beginning around maximum light, the purported \siii{} profile may in fact be heavily blended with, or indeed dominated by, \ion{Fe}{ii} \citep{obs--2011ay}. As demonstrated by Fig.~\ref{fig:12z_comp}, there are some differences between our models and the observed \siii{} profiles, which may be caused by the lack of \ion{Fe}{ii} features produced in the models.

\par

\begin{figure*}
\centering
\includegraphics[width=12.5cm]{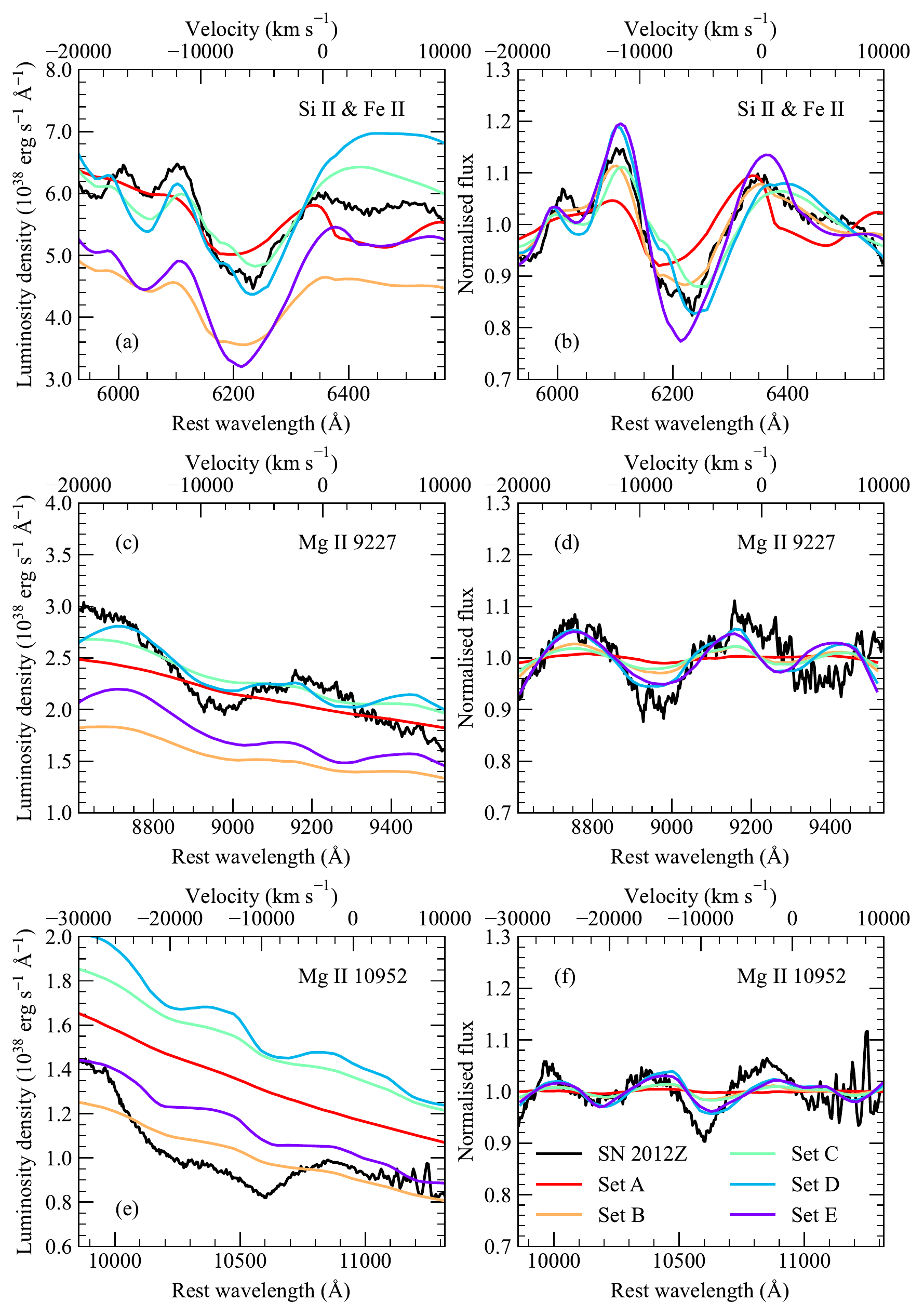}
\caption{Comparison of SN~2012Z to models with different plasma states in the model ejecta. \textit{Left: }Models and spectra are shown in absolute luminosity, with no scaling applied. \textit{Right: }All spectra are normalised by fitting the region shown for a continuum that is then divided out. The continuum is removed to demonstrate the shapes of the features and due to the limitations of \textsc{tardis} in producing accurate flux levels across all wavelengths at these times. In \textit{Panels (a) \& (b)}, velocities are given relative to \siii{}, while other panels show velocities relative to the specified \ion{Mg}{ii} line.
}
\label{fig:blend_comp}
\centering
\end{figure*}

\begin{table}
\centering
\caption{\textsc{tardis} model parameters and properties.}\tabularnewline
\label{tab:model-params-blend}\tabularnewline
\resizebox{\columnwidth}{!}{
\begin{tabular}{ccccc}
\hline
\hline
Spectrum & Time since       & Luminosity           & Inner boundary         & Blackbody    \tabularnewline
     & explosion (days) & ($\log$~L$_{\odot}$)  & velocity (km~s$^{-1}$) & temperature (K)  \tabularnewline
\hline
\hline
\multicolumn{5}{c}{Set A - Reference} \tabularnewline
\hline
NIR     & 14.6 & 9.05 & 9\,400 & 9\,000 \tabularnewline
Optical & 16.8 & 9.04 & 9\,200 & 8\,400 \tabularnewline
\hline
\multicolumn{5}{c}{Set B - Reduced temperature} \tabularnewline
\hline
NIR     & 14.6 & 8.78 & 9\,400 & 7\,800 \tabularnewline
Optical & 16.8 & 8.90 & 9\,200 & 7\,800 \tabularnewline
\hline
\multicolumn{5}{c}{Set C - Increased time} \tabularnewline
\hline
NIR     & 19.0 & 8.93 & 9\,400 & 7\,300 \tabularnewline
Optical & 21.8 & 9.03 & 9\,200 & 7\,200 \tabularnewline
\hline
\multicolumn{5}{c}{Set D - Increased density} \tabularnewline
\hline
NIR     & 19.0 & 8.91 & 9\,400 & 7\,500 \tabularnewline
Optical & 21.8 & 9.03 & 9\,200 & 7\,500  \tabularnewline
\hline
\multicolumn{5}{c}{Set E - Reduced velocity} \tabularnewline
\hline
NIR     & 19.0 & 8.78 & 7\,600 & 7\,800 \tabularnewline
Optical & 21.8 & 8.92 & 7\,500 & 7\,800  \tabularnewline
\hline
\hline
\multicolumn{5}{l}{} \tabularnewline
\end{tabular}
}
\end{table}

One of the primary assumptions of \textsc{tardis} is that of a sharp photosphere separating optically-thick and optically-thin regions that does not vary with wavelength. In practice, the lower opacity at NIR wavelengths should result in a photosphere that is deeper inside the ejecta than for optical wavelengths and hence NIR packets should also be injected deeper inside the model. This assumption means that \textsc{tardis} is generally ill-suited for calculating consistent optical and NIR spectra. The flux at longer wavelengths is typically over-predicted due to packets being injected too close to the ejecta surface, which also leads to shorter path lengths, and hence optical depths, for these longer wavelength packets. In spite of this, \textsc{tardis} is still able to calculate accurate excitation and ionisation populations, and opacities, for these regimes as these properties do not require information on the location of the photosphere. We therefore look to answer the question of which physical conditions (ionisation state(s), excitation, temperature) could better reproduce both the \siii{} and \mgiit{} profiles. Due to the strong blending of \mgiit{} with another feature around $\sim$10\,200~\AA, we also investigate the \ion{Mg}{ii}~$\lambda$9\,227 feature.

\begin{figure*}
    \centering
    \begin{subfigure}[b]{0.49\textwidth}
        \includegraphics[width=8.6cm]{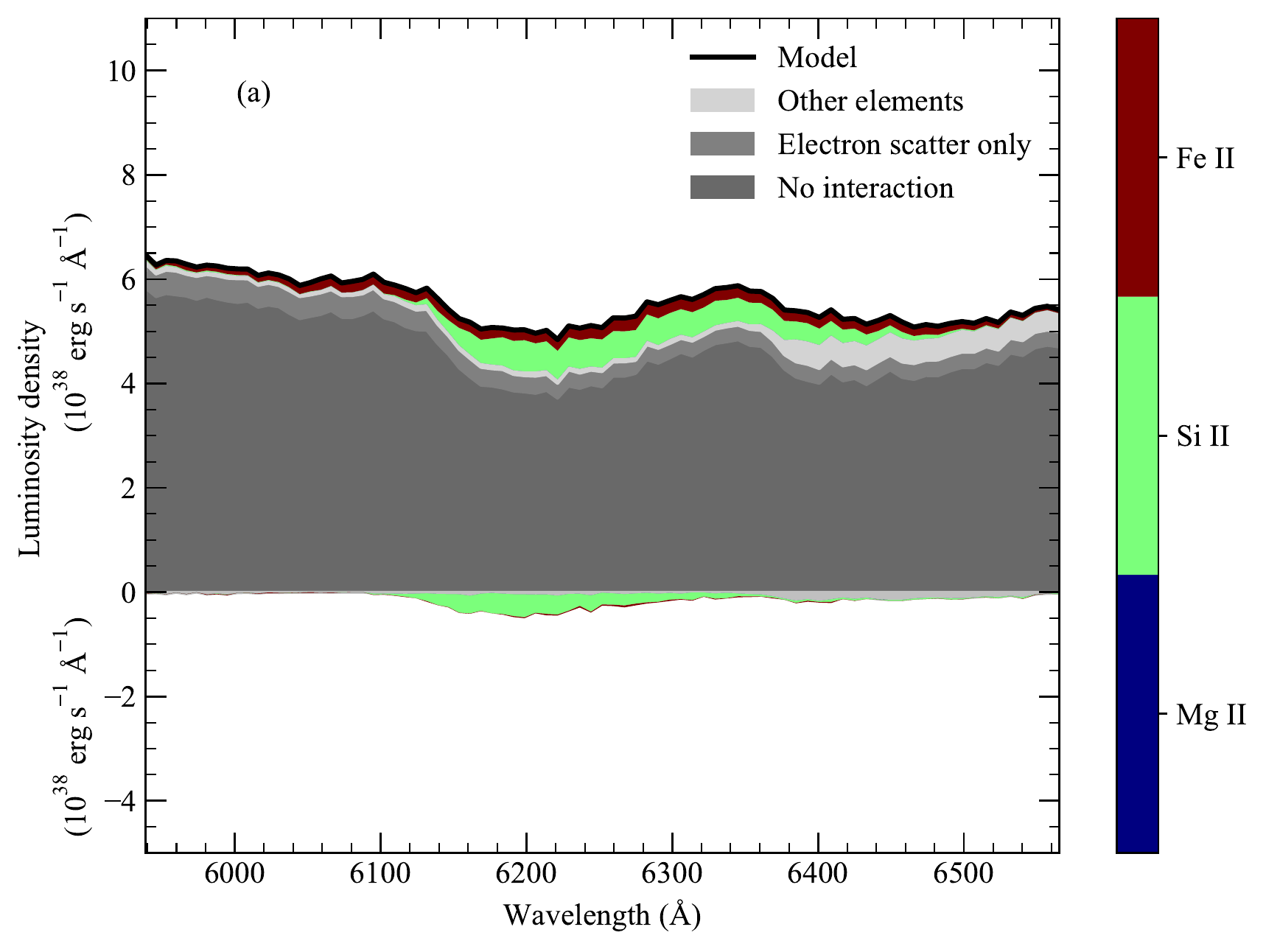}
    \end{subfigure}
    \begin{subfigure}[b]{0.49\textwidth}
        \includegraphics[width=8.6cm]{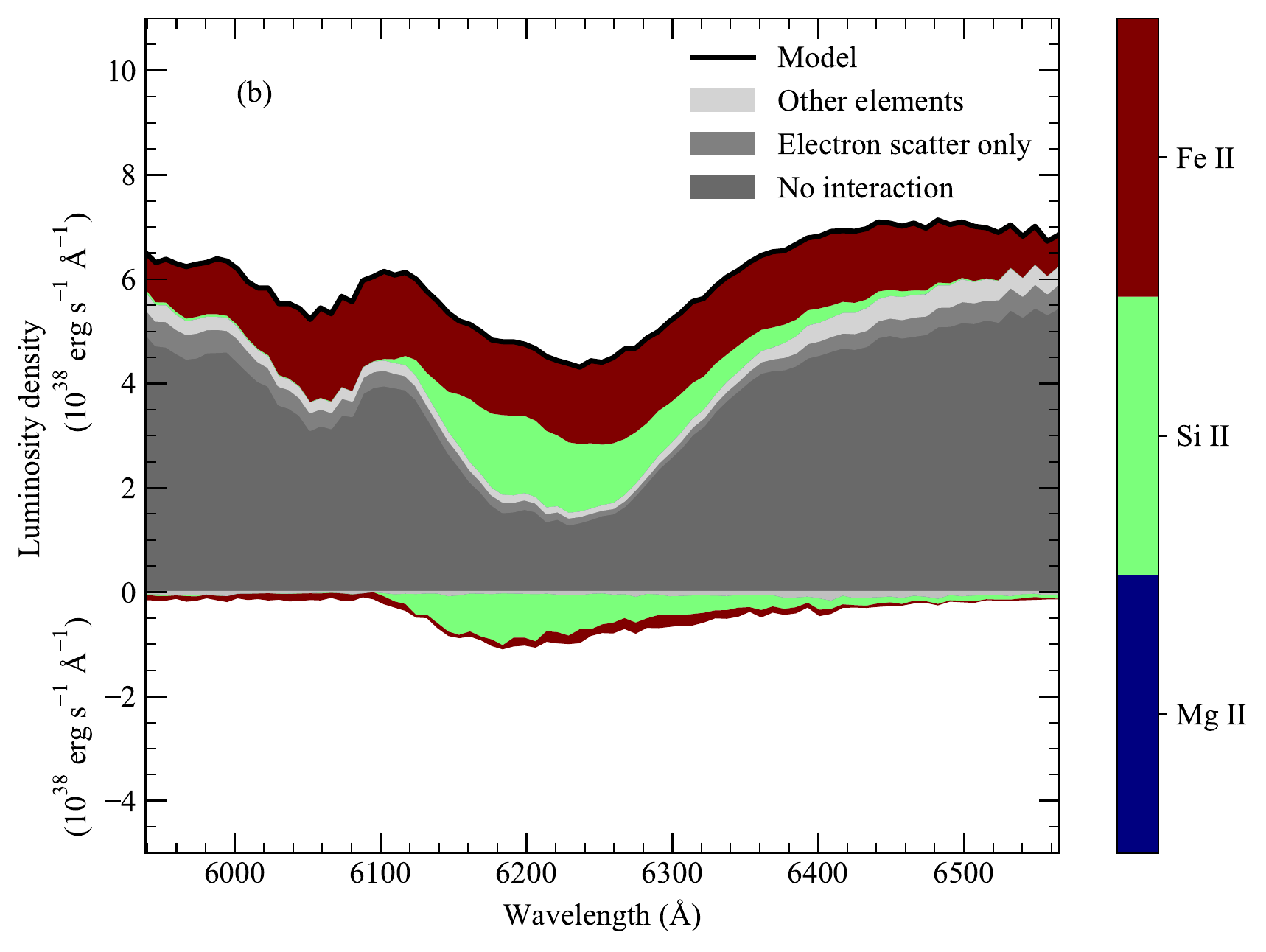}
    \end{subfigure} 
    \\
    \begin{subfigure}[b]{0.49\textwidth}
        \includegraphics[width=8.6cm]{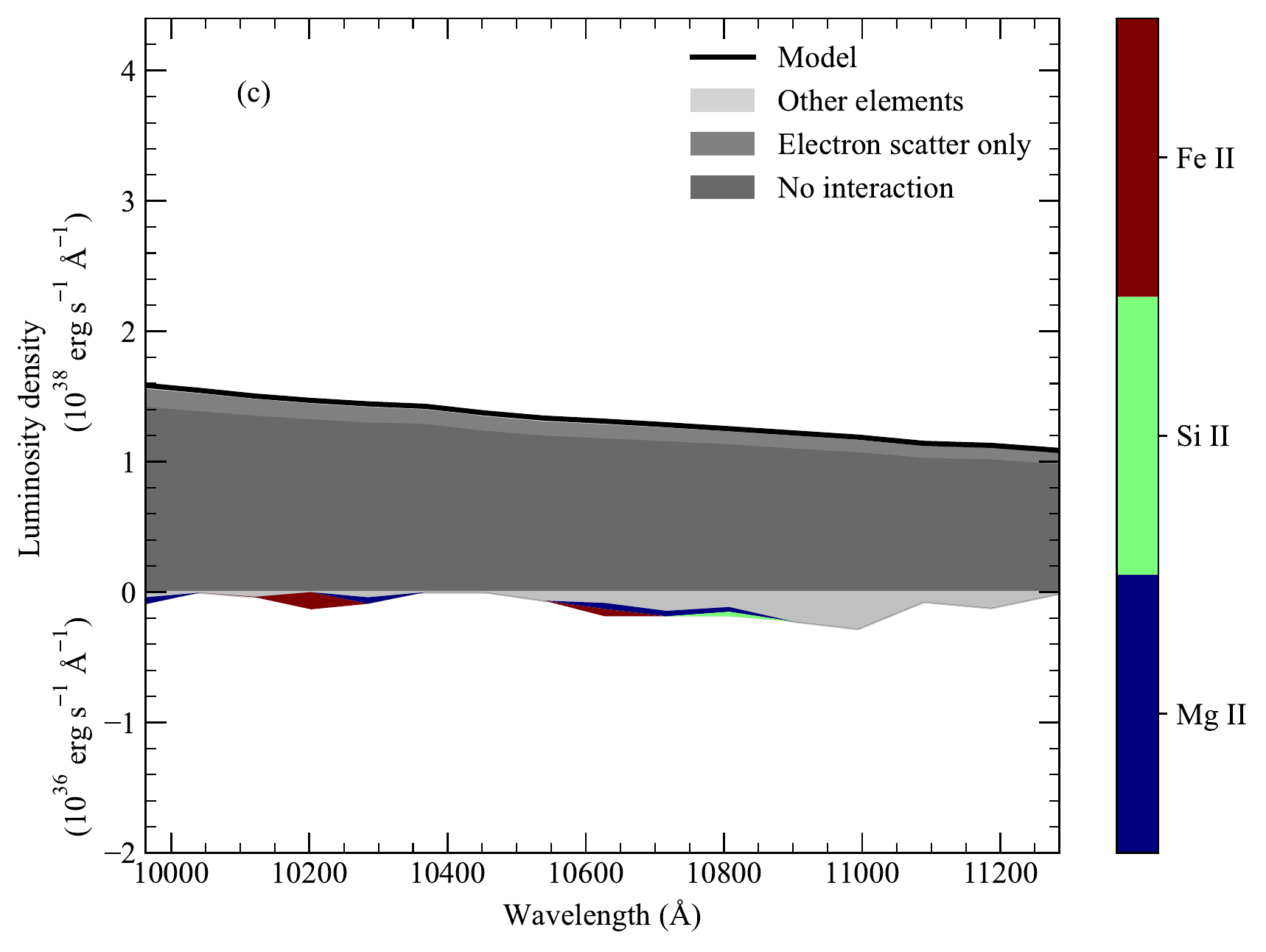}
    \end{subfigure}
    \begin{subfigure}[b]{0.49\textwidth}
        \includegraphics[width=8.6cm]{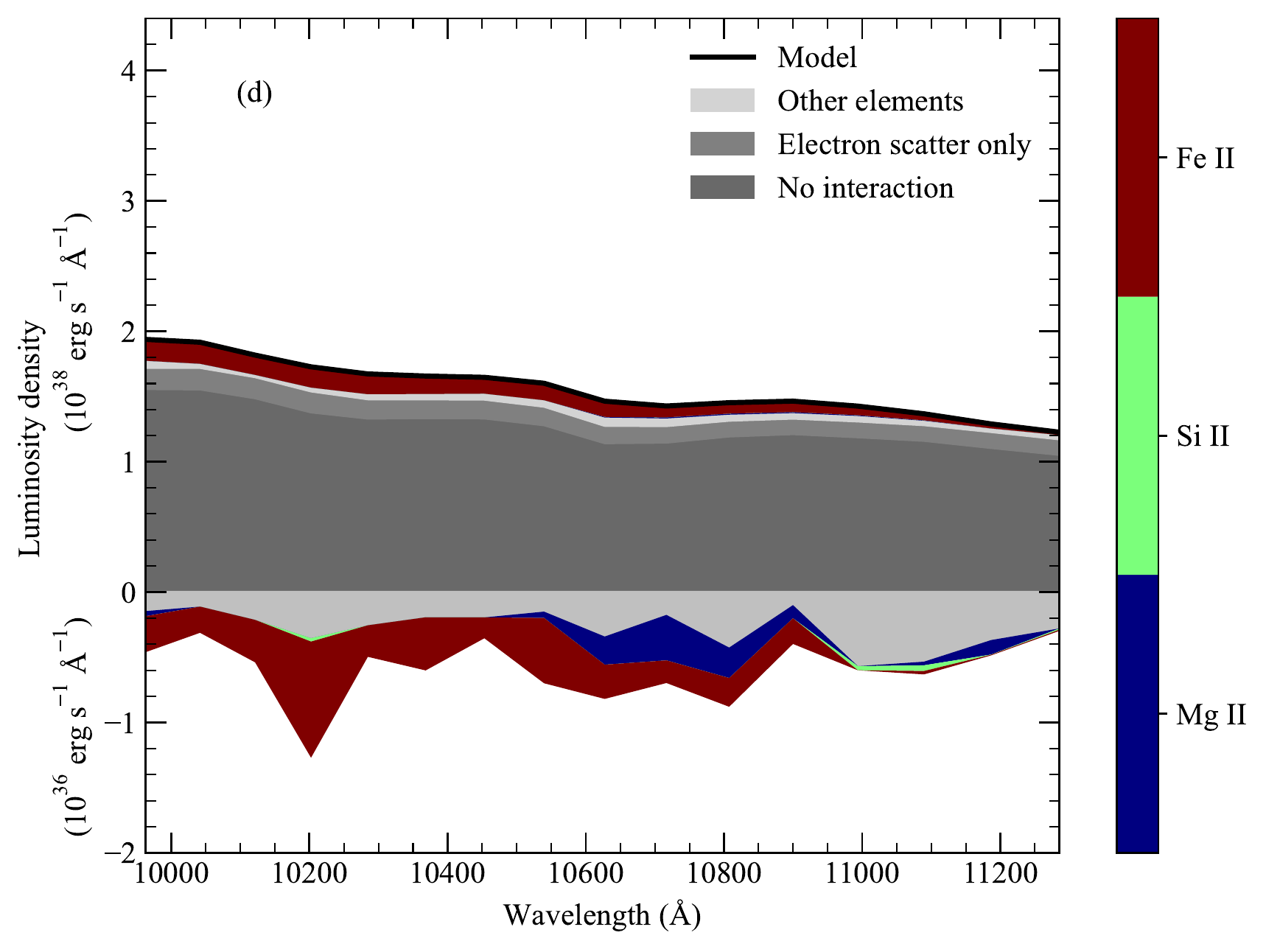}
    \end{subfigure}
    \caption{Contribution of specific ions to our model spectra exploring different plasma states. \textit{Left} panels show models from set A, while \textit{Right} panels show models from set D. As in Fig.~\ref{fig:kromer}, we colour code histograms based on the species with which escaping Monte Carlo packets experienced their last interaction.
   }
    \label{fig:blending_kromer}
\end{figure*}

\par

We present five sets of models that each explore different ejecta conditions and demonstrate their impact on the model spectra. The input parameters used for each model are given in Table~\ref{tab:model-params-blend}, but we stress that these values are given for reference only. These models are not intended to reproduce the entire spectra, but to investigate what plasma states are required for the features of interest. The first set (A) is based on the mixing models presented thus far. The optical spectrum for this set has already been discussed in Sect.~\ref{sect:models}. Here, we also add a model with suitable parameters for the NIR spectrum, which is approximately two days earlier than the optical spectrum at $+1.8$\,d. As previously mentioned, there are some disagreements between these models and observations of SN~2012Z, therefore they are provided as the reference points through which we can determine the impact of various input parameters. In set B, we simply lower the temperature of the input black body. For our remaining three model sets, we arbitrarily increase the time since explosion by 30\%. This is done to increase the path lengths packets must travel before exiting the ejecta. The input parameters of \textsc{tardis} are at least somewhat degenerate, therefore we find similar results for combinations of different explosion times and other input parameters compared to those given in Table~\ref{tab:model-params-blend}. Hence the exact values are unimportant and we again note that the purpose of this exercise is to explore the ionisation and excitation state of the ejecta required to better match the observations and somewhat circumvent the limitations of \textsc{tardis} when modelling longer wavelengths at relatively late times. For model set C, we use the same inner boundary velocity as in set A. The increased time since explosion has also resulted in a drop in density of the ejecta. Therefore, in set D, we increase the mass throughout the ejecta by a constant factor (1.3$^3$) to account for the reduction in density due to the 30\% increase in radius. Alternatively, in set E, we allow the inner boundary velocity of the models to recede deeper inside the ejecta, such that the density at the photosphere is the same as in sets A \& B. For all models, we use the same underlying density profile and composition, namely the angle-averaged N10def model.

\par

In Fig.~\ref{fig:blend_comp} we show a comparison between each of the model spectra and SN~2012Z. Figure~\ref{fig:blend_comp} also shows the model and observed spectra in normalised flux, where the continuum in the regions shown has been divided out. The purpose of these normalised comparisons is to demonstrate the shapes of spectral features produced. As mentioned, we expect that the flux levels produced by \textsc{tardis} will be affected by the photospheric assumption, while the features produced will be accurate. Figure~\ref{fig:blend_comp}(a) shows that our standard mixing model (set A) predicts a \siii{} feature that appears to be at somewhat higher velocities than the data and does not reproduce the \feiit{} feature around $\sim$6\,100~\AA. This is also shown by Fig.~\ref{fig:blending_kromer}, which shows the contribution of individual ions. As demonstrated by Fig.~\ref{fig:blending_kromer}(a), this region is clearly dominated by \ion{Si}{ii}. All other models however, produce features that are much more similar to SN~2012Z, with either a broader, flat-bottomed absorption trough or two clearly discernible profiles, due to the presence of \feiif{}. This is shown for set D in Fig.~\ref{fig:blending_kromer}(b), which demonstrates the increased absorption and fluorescence from \ion{Fe}{ii}. The luminosities of sets B \& E are clearly somewhat lower (by $\sim$30\%; Fig.~\ref{fig:blend_comp}(a)) than SN~2012Z, but the shapes of the profiles are comparable (Fig.~\ref{fig:blend_comp}(b)). In addition to the increased \feiif{} absorption, all models also show much stronger \feiit{} features (see Fig.~\ref{fig:blending_kromer}(b)), consistent with what is seen in SN~2012Z. These models further demonstrate that, if there is considerable \feiit{} absorption in SN~2012Z, the lack of high velocity \siii{} in the maximum light spectrum could simply be due to the presence of an additional absorption profile cutting short the \siii{} blue-wing.

\par

The \mgiin{} profiles are shown in Figs.~\ref{fig:blend_comp}(c) \& (d). It is clear that the standard mixing model from set A does not reproduce the observations at these wavelengths. The NIR spectrum from set A lacks any strong absorption as a result of the photospheric approximation in \textsc{tardis}. As with the optical \siii{} and \feiif{} features however, all other model sets produce improved agreement. Indeed, these models show that the absorption minima and widths are comparable to SN~2012Z, but the luminosity of sets B \& E are again somewhat lower. Finally, the \mgiit{} profiles are shown in Figs.~\ref{fig:blend_comp}(e) \& (f). Again we find that model set A lacks any strong absorption. Figure~\ref{fig:blending_kromer}(c) shows that, for this model, essentially all packets escaping at these wavelengths do not interact with the ejecta. Similar to previous comparisons, model sets B -- E produce features much more similar to SN~2012Z in terms of their shapes, strengths, and velocities. With the exception of model set B, the luminosities of all spectra are higher than in SN~2012Z. Again, this is likely a result of the photospheric approximation in \textsc{tardis}. Figure~\ref{fig:blending_kromer}(d) shows the increased \mgiit{} absorption for set D, although there is also a contribution from \ion{Fe}{ii}. The \mgiit{} profile in SN~2012Z is clearly complicated. Even compared to the \mgiin{} profile, this feature shows a much sharper absorption minimum, while the feature at $\sim$10\,200~\AA\, clearly impacts the blue-wing. Fig~\ref{fig:blending_kromer}(d) indicates this feature is likely due to \ion{Fe}{ii}~$\lambda$10\,504. As demonstrated by Figs.~\ref{fig:blend_comp}(e) \& (f), the models do however show the location of the \mgiit{} absorption minima are comparable to SN~2012Z, but they are not as sharp as observed.

\par

The differences between the observed \mgiin{} and \mgiit{} profiles complicate the interpretation as the \mgiit{} feature is clearly quite complex. Nevertheless, model sets B -- E produce features comparable to both \ion{Mg}{ii} profiles and the \siii{} \& \feiif{} blend. We again note that these models use the angle-averaged composition from the N10def explosion model, which is well-mixed (see Fig.~\ref{fig:model_composition}). Therefore, based on the agreement between the \siii{} \& \feiif{} and \mgiin{} profiles, there is no indication from model sets B -- E that stratification of silicon relative to magnesium is required for SN~2012Z, provided these or similar ejecta conditions could be reproduced.

\begin{figure}
\centering
\includegraphics[width=\columnwidth]{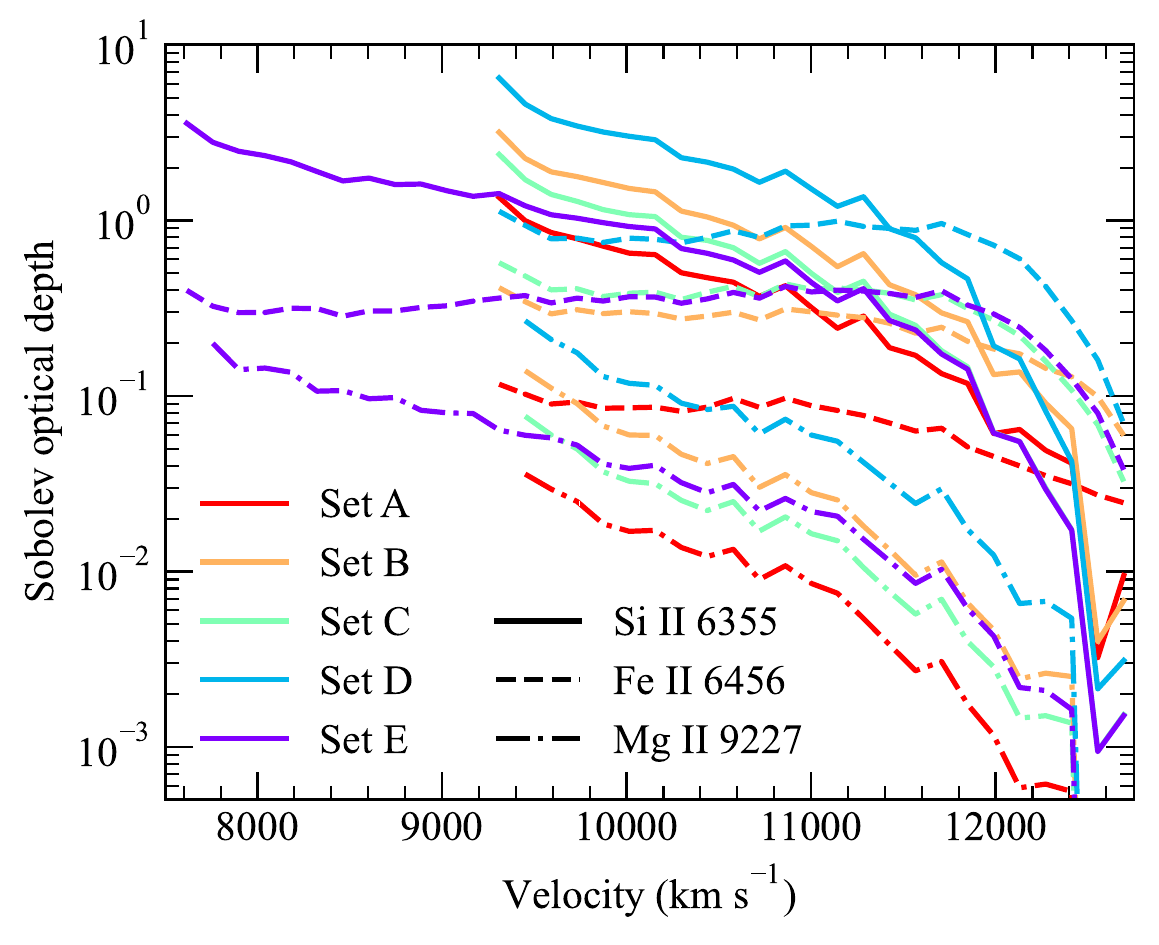}
\caption{Sobolev optical depths calculated by \textsc{tardis} for our model sets exploring different plasma states. 
}
\label{fig:blend_opt_depth}
\centering
\end{figure}

\begin{figure*}
\centering
\includegraphics[width=\textwidth]{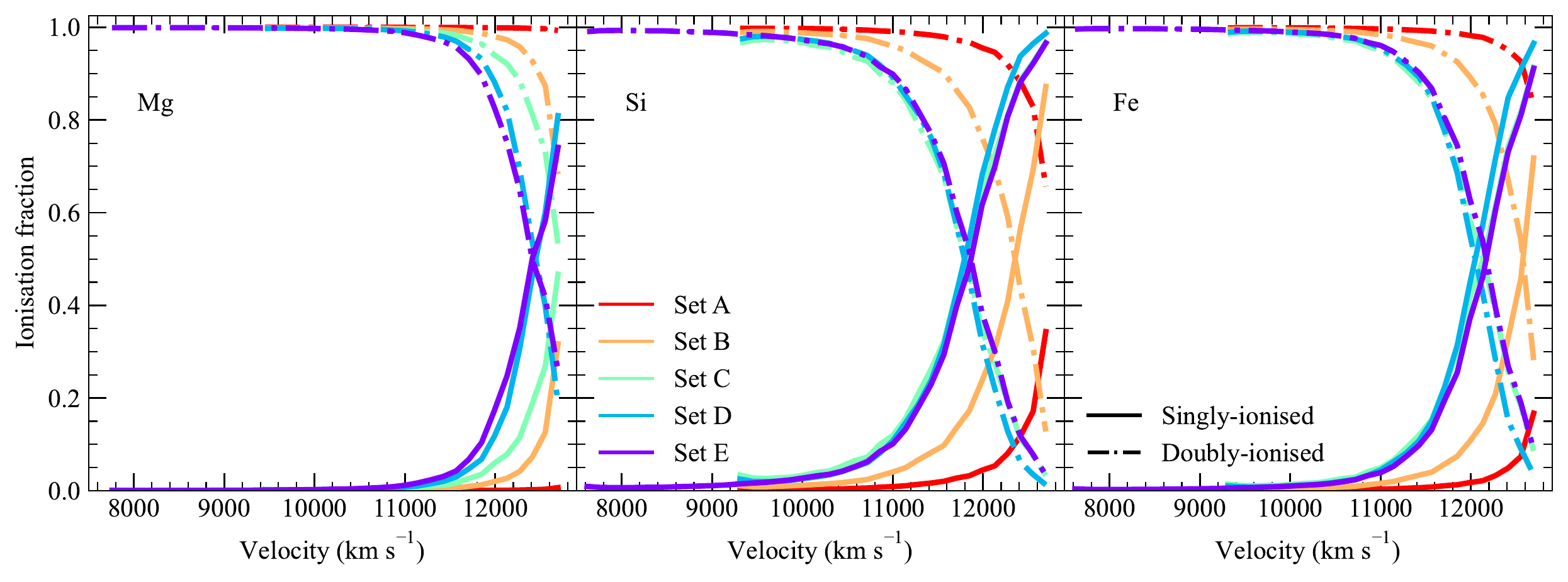}
\caption{Ionisation fraction of Mg, Si, and Fe for models within each of our sets exploring different plasma states. Solid lines show the fraction of each species that is single-ionised, while dash-dot lines show the fraction that is doubly-ionised.
}
\label{fig:blend_ions}
\centering
\end{figure*}

\par

In Fig.~\ref{fig:blend_opt_depth} we show the Sobolev optical depths of \siii{}, \feiif{}, and \mgiin{} for each of the models. The optical depth of \mgiit{} follows the same trend as \mgiin{} for all models, although is $\sim2\times$ lower, and is omitted here for clarity. The lack of significant contribution from \feiif{} to blending in the 6\,200~\AA\, feature for model set A is clearly apparent, due to the significantly smaller optical depth of \feiif{} compared to \siii{} at all velocities except the very outermost ejecta. In contrast, for model sets B -- E, the optical depth of \feiif{} is comparable to \siii{}, or even larger in some regions of the outer ejecta. Therefore, our models indicate that to produce strong \feiif{} absorption such that there is significant blending with \siii{}, a typical Sobolev optical depth of $\gtrsim$0.3, and within a factor of a few of \siii{}, would be required. Aside from changes in the plasma state, an increased optical depth could also be achieved through an increased iron abundance. As mentioned, our models use the composition of the N10def explosion model, therefore it is plausible that increasing the iron abundance relative to this model may also produce strong iron blending. Alternatively, it is also possible that the modest $\sim3\times$ increase in Sobolev optical death required to improve agreement relative to model set A is within the uncertainty of the ionisation and excitation approximations used by \textsc{tardis}, which could simply be offset by small amounts.

\par

One of the key differences between model sets B -- E compared to the reference set A is the reduction in temperature. The temperature of the black body at the photosphere is reduced by 600~--~1\,200~K for optical models and 1\,200~--~1\,700~K for NIR models (see Table~\ref{tab:model-params-blend}). In Fig.~\ref{fig:blend_ions} we show the fraction of magnesium, silicon, and iron that is either singly- or doubly-ionised for each model. The lack of strong \ion{Mg}{ii} or \ion{Fe}{ii} features in set A is likely due to the fact that the ejecta is predominantly doubly-ionised for these two species. Indeed, magnesium is approximately 100\% doubly-ionised throughout the entire model ejecta. Conversely, in model set E, for example, there is a significant increase in the \ion{Mg}{ii} fraction towards the outer ejecta, $\sim$25\% at 12\,000~\kms{}. Unsurprisingly, the strengths of the features shown in Fig.~\ref{fig:blend_comp} are strongly correlated with the increase in the singly-ionised population. Interestingly, we find that magnesium is more highly ionised than silicon in all model sets, including set A. For example, in set D, at 12\,000~\kms{}, magnesium is $\sim$90\% \ion{Mg}{iii} while silicon is only $\sim$40\% \ion{Si}{iii}. This clearly demonstrates the possibility of stratification in ionisation rather than abundance. In other words, differences between blue-wings of different elements may simply reflect their different ionisation states, rather than the lack of one element at a given velocity.

\par

In summary, we have presented a series of models exploring different plasma states for the ejecta. Based on these models, we find that the different \siii{} and \mgiit{} (or \mgiin{}) profiles could be reproduced in a well-mixed model due to strong blending with \ion{Fe}{ii}, consistent with previous analysis by \cite{obs--2011ay}. In order to achieve significant blending of the 6\,200~\AA\, absorption feature, our models indicate that a Sobolev optical depth for \feiif{} of $\gtrsim$0.3 is required. If the necessary plasma state cannot be naturally achieved in a well-mixed model, stratification could provide an alternative explanation. Such stratification however could be in either abundance, as suggested by \cite{comp--obs--12z}, or in ionisation and excitation state. For all of our models, we find that magnesium is more highly ionised than silicon throughout. Therefore, any difference in line profiles does not necessarily require separation of elements in physical space. The relevant ions may simply not exist in the same regions and hence give the appearance of a physical separation for given elements. Our models indicate that this can be naturally achieved even in a well-mixed model.

\section{Discussion}
\label{sect:discussion}

Our models show that the spectroscopic features of SN~2012Z are consistent with a well-mixed ejecta structure. Based on our models, there is no indication that a stratified distribution is required to match the observations. Stratification could be allowed, however stratification in ionisation state, rather than abundance, is equally plausible and can be achieved even in a well-mixed model. Alternatively, we find that blending also provides a natural explanation for the observed differences in line profiles. Finally, a steeper density profile, relative to the pure deflagration models of \cite{fink-2014}, provides another possibility for improving agreement with the observations.

\par

\subsection{Presence and distribution of carbon}
\label{sect:carbon}

Based on our analysis, a well-mixed or near-uniform composition provides the best agreement with observations of SN~2012Z up to shortly after maximum light. One of the challenges for these models however, is the strong carbon features produced at early times, which are much weaker in the observed spectra. \cite{12bwh} previously argued that a reduced carbon abundance, relative to the pure deflagration models of \cite{fink-2014}, can provide improved agreement with the SNe~Iax SN~2005hk and PS1-12bwh. A similar change in the carbon mass fraction here would likely improve agreement with the observations.

\par

The best matching distributions of carbon for our models are in contradiction to the model that \citet{barna--18} propose for SN~2012Z. In this model, the carbon fraction in the outer ejecta steadily increases (up to 0.10 -- 0.25 between 12\,000 -- 16\,000~\kms{}) in the outermost regions. These values are generally higher than the model we present here, which already produces carbon features that are too strong at early times. The cause of this discrepancy is likely due to the density profile used by \cite{barna--18}. As shown in Fig.~\ref{fig:dens}, the density is generally lower in the outer ejecta, but also extends to significantly higher velocities, relative to the \cite{fink-2014} models. \cite{barna--18} state that the carbon fractions predicted in these regions should be considered upper limits. Similar results are also found by \cite{barna--21} in the analysis of SN~2019muj -- the sharp cut-off in the assumed density profiles means that the abundances in these outermost regions (i.e. where the carbon appears in their model) are highly uncertain. Considering these points together, the upper limit on the carbon abundance in the outermost ejecta is likely lower (than $\sim$10\%), given that our models also predict strong carbon features and should be more sensitive to these regions (due to the higher density; see Fig.~\ref{fig:dens}). Based on previous work by \cite{12bwh}, a limit of a few percent may be more appropriate, although again this may be quite uncertain depending on the density profile and data available.

\par

Apart from the outermost regions, \cite{barna--18} also argue that carbon is excluded from the inner ejecta ($\lesssim$11\,000~\kms{}) of most SNe~Iax. Similar results are also found by \cite{barna--21} for SN~2019muj. In this case, the significantly lower velocities of SN~2019muj mean that carbon does not appear below 6\,500~\kms{} in the model. Again, our results are in contradiction to these claims. The carbon features predicted by our well-mixed models around and after maximum light are weaker than those at early times and in closer agreement with the observations. This is also demonstrated by \cite{kromer-13} for the case of the N5def model and SN~2005hk -- despite a near-uniform carbon abundance there are no strong carbon features predicted shortly after maximum light. \cite{barna--18} (and indeed \citealt{barna--21}) only consider two possibilities for carbon in the inner regions -- a uniform carbon abundance comparable to the \cite{fink-2014} predictions or zero carbon. They do not attempt to place estimates on the amount of carbon that could be included in these regions, but simply not produce features. Therefore the uniform carbon abundance is again best considered as an upper limit, while the true carbon fraction may be a few percent or less. In this case, it may not be possible to distinguish this from an ejecta with no carbon in the centre.

\par

The distribution of carbon is of particular importance as \citet{barna--18} argue that the gradient for the carbon distribution found from their work is inconsistent with predictions from pure deflagration models. As discussed, our models contradict this claim. Our models do not show strong carbon features after maximum light and therefore we find no evidence that a carbon gradient is required. Simply reducing the carbon abundance throughout the ejecta may also produce favourable agreement.

\par

Our models and the N5def model of \citet{kromer-13} do, however, predict strong carbon features before maximum light that are not observed. This is most apparent for our earliest model spectrum (Fig.~\ref{fig:12z_comp}(a)), which shows clear features due to \ion{C}{ii}~$\lambda$4\,270, $\lambda$4\,740, and $\lambda$6\,580 that are either much weaker or not visible in SN~2012Z. While carbon features have been identified, or tentatively identified, in some SNe~Iax \citep{2005hk--spec--pol, early--late--08ha--obs, obs--07qd, foley--13, tomasella--2016} and \cite{foley--13} claim that $\gtrsim$80\% of SNe~Iax with spectra before maximum light show signs of carbon, the features are generally much weaker than those predicted by our models. Therefore it is likely that the carbon fraction in the outer ejecta is too high in these models, rather than in the inner ejecta.

\par

Carbon itself is not produced during the explosion, but the amount of carbon contained within the ejecta is a result of unburned material entrained during the turbulent flame propagation. Different explosion models can produce different total carbon masses within the ejecta due to this entrainment. Indeed, pure deflagration models presented by \citet{jordan--12} show the percentage of unburned material (carbon and oxygen) within the ejecta to be $\sim$43 -- 57\%, while models with similar ejecta masses presented by \cite{fink-2014} contain $\sim$28 -- 37\% unburned material. Our model spectra indicate that a lower carbon fraction in the outer ejecta likely could provide improved agreement, similar to what was found by \citet{12bwh} for SN~2005hk and PS1-12bwh. We note however that \citet{12bwh} used a factor of ten reduction in the carbon abundance, but did not explore the range of carbon fractions allowed. It is unclear whether a factor of ten reduction could be achieved through differences in entrainment alone or if a smaller reduction is also allowed by the data. 

\par 

Our models show that a reduced carbon fraction in the outer ejecta, relative to the \cite{fink-2014} models, is preferred by the data. As the post-maximum spectra do not predict strong carbon features, we are unable to constrain the carbon distribution further. In other words, we are unable to determine whether the carbon fraction should be reduced throughout the ejecta or in only the outermost regions. That the carbon fraction of the \cite{fink-2014} models appears somewhat too high relative to observations of SNe~Iax is itself not an argument against mixing or indeed layering. The only concrete statement that can be made is the carbon fraction needs to be reduced somewhere in the ejecta, but the observations are insensitive to whether or not this implies layering, as the models are insensitive to a reduction in only the outer ejecta compared to a reduction by a constant amount throughout the entire ejecta. Future modelling is required to investigate which of these possibilities is more likely and preferred, however it is clear that with the data currently available one cannot make definitive statements that carbon stratification is necessary. Alternatively, a steeper density profile may also help to reduce to carbon mass in the outer ejecta (see Sect.~\ref{sect:density}).

\subsection{Density profile}
\label{sect:density}

Our models that assume angle-averaged compositions from pure deflagration simulations show good agreement with SNe~Iax (see Sect.~\ref{sect:model_spectra_comp}, Fig.~\ref{fig:12z_comp}), except for the strong \ion{C}{ii} features predicted by the models at early phases. We do find however, that removing the high velocity ejecta entirely is one method to produce spectral features more similar to SN~2012Z. Therefore, this would indicate that rather than stratification, a steeper density profile is a possibility. This could also be one alternative to producing a reduced carbon mass in the outer ejecta, which appears to be preferred by observations of SN~2012Z. In addition, \citet{barna--18} argue for a decreasing mass fraction towards the outer ejecta for all elements (except oxygen, which acts as a filler), but a similar effect may be achieved with a steeper density profile.

\par

A steep density profile was used by \citet{05hk--400days} for modelling the light curve and spectra of SN~2005hk. Relative to the W7 model \citep{nomoto-w7}, the E03 model presented by \citet{05hk--400days} had a kinetic energy scaled down by a factor of $\sim$4. Although the E03 model has a lower kinetic energy than the W7 model, it is a factor of $\sim$2 higher than the N5def pure deflagration model presented by \citet{kromer-13}, which also shows good agreement with SN~2005hk. Both models extend to a maximum velocity of $\sim$12\,000~km~s$^{-1}$, while the ejecta mass in the E03 model is $\sim$3.8 times higher. These density profiles are also shown in Fig.~\ref{fig:dens}, along with the model used by \cite{barna--18}.

\par

 Previous studies have already demonstrated how the ejecta mass predicted by the pure deflagration models of \citet{fink-2014} is lower than observed in SNe~Iax and as a result the models evolve too quickly \citep{kromer-13, 15h}; hence there is scope for improved agreement with alternative density structures. We also note that the high ejecta density in the E03 model meant that [\ion{O}{i}] was not observed at late times, despite the high oxygen mass in the ejecta. The lack of strong [\ion{O}{i}] features is also consistent with observations of SN~2005hk, further indicating a steep density profile (or a high density at low velocities) could be preferred.

\par

We speculate that, relative to the pure deflagration models of \citet{fink-2014}, a steeper density profile and an increased ejecta mass could overcome some of the apparent disagreements between similar pure deflagration models and SNe~Iax. We note that, as shown in Fig.~\ref{fig:dens}, the density profiles used in this work are already shallower than others used in the literature. The increased ejecta mass should not come from an increase in unburned material contained within the ejecta, as the deflagration models already show \ion{C}{ii} features that are too strong. Instead, more of the progenitor white dwarf should be burned to IMEs or IGEs. In the case of an increased IGE abundance, this would also lead to an increased brightness (due to the higher $^{56}$Ni mass). Pure deflagrations can already produce a range of $^{56}$Ni masses, therefore whether this simply results in all models shifting to higher luminosities or if this can be counteracted by a change in other parameters (such as central density, ignition location, metallicity, etc.) remains to be seen. Based on their pure deflagration models, \cite{long--14} demonstrate the significant impact the initial conditions of the explosion can have on the range of ejecta produced, including a variety of different density profile shapes and some with steep drop-offs in the outer ejecta. Future explosion models should further explore the range of ejecta structures that could be produced and whether this would indeed agree better with observations of SNe~Iax. Caution must be applied however when using steep density profiles to infer the composition of the ejecta, due to the inherent degeneracy between the shape of the density profile and composition. Spectra as early as possible (within a few days of explosion) would be required to place firm constraints on the outer regions in this case.

\subsection{Pulsational delayed detonations versus pure deflagrations}

Based on our analysis, SN~2012Z is consistent with a well-mixed ejecta and does not appear to require stratification of specific elements. This does not in itself however indicate that SN~2012Z must also be consistent with any existing explosion scenario or literature model. Here, we briefly discuss the merits of two competing explosion scenarios, pulsational delayed detonations and pure deflagrations, but we also note the possibility that neither scenario is the `correct' one.

\par

\citet{comp--obs--12z} suggest that SN~2012Z (and, by extension, other SNe~Iax) is consistent with a PDD explosion model, such as those presented by \citet{hoeflich-02} and \citet{dessart--2014b}. No direct comparison between SN~2012Z and photospheric phase spectra predicted by PDD models has been performed in the literature to date, therefore we are unable to determine whether our models represent improved agreement. It is also unclear whether the carbon distribution or density profile predicted by the PDD scenario is preferred over the pure deflagration scenario. Nevertheless, some models presented by \citet{hoeflich-02} do show a small carbon fraction in the outer ejecta and may produce lower carbon masses than predicted by pure deflagrations.

\par

One of the claims in favour of the PDD scenario suggested by \citet{comp--obs--12z} is the measured blue-wing velocities of \siii{} and \ion{Mg}{ii}~$\lambda$10\,952. The blue-wings of these features are argued to be offset in velocity-space, therefore these elements do not extend to cover the same region of the ejecta. Such a layered structure would be more similar to the PDD models of \citet{hoeflich-02} and would not be consistent with pure deflagration models, where silicon and magnesium are always co-spatial with each other. Based on our models, the blue-wing velocity measured from individual absorption profiles does not necessarily correspond to the maximum velocity at which an ion is present. For SNe~Iax in particular, the significant degree of blending means that absorption profiles can not necessarily be used to diagnose the velocity extent of individual elements and strong blending could significantly alter the shapes of line profiles for different elements, complicating the interpretation of velocity measurements. Our models also show that stratification of ionisation state is naturally reproduced even for well-mixed models, therefore differences in the maximum velocities of specific ions may not imply stratification in abundance. Furthermore, our models indicate that a steeper density profile provides better agreement with SN~2012Z than a stratified silicon abundance.

\par

Aside from the blue-wing velocities of \siii{} and \ion{Mg}{ii}~$\lambda$10\,952, \citet{comp--obs--12z} also favour a Chandrasekhar-mass explosion based on the emission profile of the [\ion{Fe}{ii}]~1.64~$\mu$m feature in the late-time (+$269$\,d) NIR spectrum, which was suggested to show a `pot-bellied profile'. Similar profiles have been claimed in other SNe~Ia \citep{hoeflich--04, motohara--06} and are argued to provide evidence for a lack of $^{56}$Ni in the WD centre due to the production of stable material at high central densities (see e.g. \citealt{seitenzahl--17}). Some pure deflagration models (see e.g. figs~10~\&~11 of \citealt{fink-2014}) also produce low $^{56}$Ni fractions in the innermost ejecta and therefore may provide similar late-time spectra to PDD models. Hence this property may not be a distinguishing feature between the pure deflagration and PDD scenarios and further model spectra are required to investigate this.

\par

The presence of a bound remnant\footnote{Here, we use the term `bound remnant' to refer to the time immediately following explosion. The remnant may not necessarily remain bound at later times.} may be a key prediction of weak deflagrations. Some of the models presented by \citet{jordan--12} and \citet{fink-2014} are sufficiently low energy that they fail to completely unbind the exploding white dwarf, leaving behind a (potentially massive) bound remnant. Unfortunately, current explosion models are unable to fully resolve this remnant. If a significant fraction of the material at low velocities remains within the remnant, this may also produce an ejecta hole. Indeed, the presence of a bound remnant has been argued as a natural explanation for the appearance of SNe~Iax spectra at late-time \citep{02cx--late--spec, 05hk--400days, foley--late--iax}. Current observations of SNe~Iax suggest that even at hundreds of days after explosion, they have not entered a truly nebular phase. Spectra at these epochs still show features from low velocity, permitted iron lines \citep{02cx--late--spec, 05hk--400days, tomasella--2016}, which could be driven by a wind emanating from the remnant. A superposition of nebular and photospheric features has shown good agreement to late-time spectra \citep{05hk--400days} and the relative strengths of these features could explain the diversity observed \citep{foley--late--iax}. Such a remnant may have already been directly observed in the case of SN~2008ha, for which a faint, red source was claimed approximately four years after explosion \citep{08ha--prog}. This source was significantly brighter than expected for the SN at this phase, however, again, it cannot be ruled out that this source is a surviving companion star. We note in addition that the late-time spectra of SN~2012Z are dominated by broad emission features (rather than narrow, photospheric features) and \citet{comp--obs--12z} show that a nebular spectrum from a PDD model provides reasonable agreement with the observations. Therefore both scenarios may be broadly consistent with the observed late-time spectra of at least SN~2012Z.

%

\section{Conclusions}
\label{sect:conclusions}

In this study, we investigated the spectroscopic signatures of layering in the ejecta of SNe~Iax. As a starting point, we considered the density profiles and compositions predicted from models invoking pure deflagrations of carbon-oxygen white dwarfs and artificially stratified the ejecta. To determine the impact of mixing, we progressively smoothed the stratified ejecta using Gaussian convolution kernels of varying widths and calculated synthetic spectra using \textsc{tardis} \citep{tardis}.

\par

We compared our models with different levels of mixing to SN~2012Z, for which a layered ejecta structure has been specifically claimed, and found that heavily mixed models produced the best agreement overall, using a combination of the SNID $r$lap, redshift, and visual inspection to assess the relative level of agreement between different models. For our models with low levels of mixing, there is insufficient (or no) burned material above the photosphere for the earliest epochs, which is in disagreement with observations of SN~2012Z. In addition, we calculated model spectra using the angle-averaged compositions predicted from pure deflagration models and found that the shapes of spectral features are extremely similar to both the heavily mixed or uniform composition models and SN~2012Z. Current observations are therefore insufficient for distinguishing between these cases. It is possible that the outermost ejecta of SN~2012Z is layered to some degree, but observations approximately one week after explosion are simply insensitive to these regions. 

\par

Taking the composition predicted from pure deflagration models, we also investigated the impact of stratification on the \siii{} blue-wing by removing either the silicon abundance or entire ejecta above a range of velocities. We found that agreement with SN~2012Z was improved when neglecting the outer ejecta, thereby indicating that a steeper density profile, potentially similar to those used previously in the literature, could be preferred, rather than a stratified silicon abundance being required. 

\par

Following from previous claims by \cite{obs--2011ay}, we investigated the possibility of blending as the cause of the observed differences between the \siii{} and \mgiit{} profiles in SN~2012Z. We calculated a series of models covering different plasma states for the ejecta and found that blending can indeed provide spectral features consistent with the shapes and velocities of those in SN~2012Z. Provided the required plasma states can be achieved, such blending can be reproduced within a well-mixed model. Alternatively, all of our models indicate that magnesium is more highly ionised than silicon throughout the ejecta. Therefore, if the different profiles are not due to blending, stratification in ionisation, rather than abundance, would also provide a natural explanation even for well-mixed models. Based on an empirical analysis of the data available, our results therefore indicate that there is no evidence to suggest that the ejecta of SN~2012Z, and by extension other SNe~Iax, must be layered. We note that this does not mean the ejecta cannot be layered to some degree, but rather our main conclusion is that there is insufficient evidence that layering is a requirement or that a well-mixed composition (possibly similar to that predicted by pure deflagration models) can be excluded. 

\par

A well-mixed (but not necessarily uniform) ejecta structure is a natural consequence of the turbulent flame propagation in pure deflagration explosion scenarios. Future explosion models should investigate the extent of layering that could be realised within this scenario and the range of parameters necessary to produce density profiles in agreement with those suggested by our results. To determine the level of layering allowed by observations of SNe~Iax, a greater sample of early spectra (within days of explosion) is required.

%

\section*{Acknowledgements}

We thank the anonymous referees for their constructive comments. This work was supported by TCHPC (Research IT, Trinity College Dublin). Calculations were performed on the Kelvin cluster maintained by the Trinity Centre for High Performance Computing. This cluster was funded through grants from the Higher Education Authority, through its PRTLI program.
We are grateful for use of the computing resources from the Northern Ireland High Performance Computing (NI-HPC) service funded by EPSRC (EP/T022175).
MRM and KM are funded by the EU H2020 ERC grant no. 758638. JHG acknowledges support from a QUB ARC summer studentship. This research made use of \textsc{tardis}, a community-developed software package for spectral synthesis in supernovae
\citep{tardis}. The development of \textsc{tardis} received support from the
Google Summer of Code initiative and from ESA's Summer of Code in Space program. \textsc {tardis} makes extensive use of Astropy and PyNE.
This work made use of the Heidelberg Supernova Model Archive (HESMA), https://hesma.h-its.org.

\section*{Data Availability}
All models presented in this work are available on GitHub\footnote{\href{https://github.com/MarkMageeAstro}{https://github.com/MarkMageeAstro}}.



\bibliographystyle{mnras}
\bibliography{Magee}




\appendix


\section{Additional comparisons}
\label{apdx:comp}
\par
Here we provide additional plots showing comparisons between our mixing models and SNe 2005hk and 2014ck. These comparisons are discussed further in the text in Sect.~\ref{sect:05hk}.

\begin{figure*}
\centering
\includegraphics[width=0.94\textwidth]{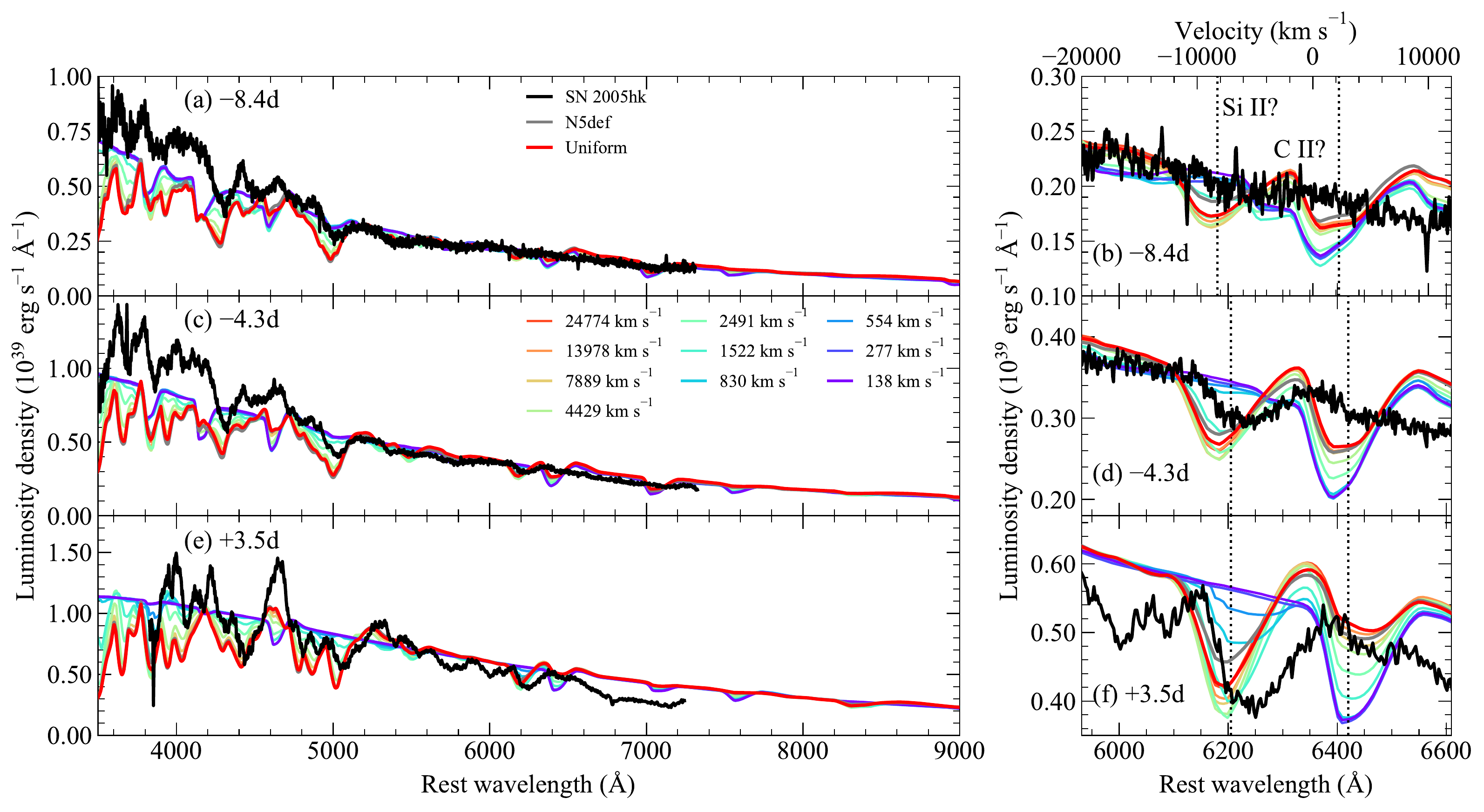}
\caption{As in Fig.~\ref{fig:12z_comp} for the case of SN~2005hk and the N5def explosion model \citep{fink-2014}.
}
\label{fig:05hk_comp}
\centering
\end{figure*}

\begin{table}
\centering
\caption{SNID $r$lap correlation coefficients for SN~2005hk.}\tabularnewline
\label{tab:snid_05hk}\tabularnewline
\resizebox{\columnwidth}{!}{
\begin{tabular}{cccccccc}\hline
\hline
Model & \multicolumn{2}{c}{2005 Nov. 02} & \multicolumn{2}{c}{2005 Nov. 06} & \multicolumn{2}{c}{2005 Nov. 14} & Mean \tabularnewline
(km~s$^{-1}$) &  & & & & &  & $r$lap \tabularnewline
\hline
\hline
& $r$lap & $z$ & $r$lap & $z$ & $r$lap & $z$ & \tabularnewline
\hline
$\phn\phn$138 & 1.46  & 0.038(0.022)  & 1.07  & 0.044(0.019)  & 0.73  & -0.009(0.000)  &  1.09  \tabularnewline
$\phn\phn$277 & 1.62  & 0.039(0.021)  & 1.03  & 0.044(0.020)  & 0.76  & -0.009(0.000)  &  1.14  \tabularnewline
$\phn\phn$554 & 1.46  & 0.038(0.022)  & 1.17  & 0.044(0.019)  & 0.74  & -0.009(0.000)  &  1.12  \tabularnewline
$\phn\phn$830 & 1.46  & 0.039(0.020)  & 1.13  & 0.043(0.020)  & 0.71  & -0.009(0.000)  &  1.10  \tabularnewline
$\phn$1\,522  & 1.50  & 0.037(0.023)  & 1.06  & 0.041(0.031)  & 0.72  & -0.009(0.000)  &  1.09  \tabularnewline
$\phn$2\,491  & 2.92  & 0.025(0.016)  & 2.43  & 0.020(0.015)  & 1.89  & 0.019(0.018)  &   2.41 \tabularnewline
$\phn$4\,429  & 5.03  & 0.022(0.009)  & 4.89  & 0.020(0.008)  & 4.96  & 0.020(0.009)  &   4.96 \tabularnewline
$\phn$7\,889  & 5.24  & 0.022(0.009)  & 5.67  & 0.020(0.007)  & 6.89  & 0.021(0.007)  &   5.93 \tabularnewline
13\,978       & 5.10  & 0.022(0.009)  & 5.68  & 0.020(0.007)  & 8.04  & 0.021(0.006)  &   6.27 \tabularnewline
24\,774       & 5.27  & 0.022(0.009)  & 5.14  & 0.020(0.008)  & 8.34  & 0.021(0.006)  &   6.25 \tabularnewline
Uniform       & 4.60  & 0.022(0.010)  & 5.37  & 0.020(0.007)  & 8.57  & 0.021(0.006)  &   6.18 \tabularnewline
N5def         & 4.86  & 0.021(0.010)  & 4.72  & 0.019(0.008)  & 7.68  & 0.021(0.007)  &   5.75 \tabularnewline
\hline
\hline
\end{tabular}
}
\end{table}

\begin{table}
\centering
\caption{SNID $r$lap correlation coefficients for SN~2005hk surrounding the \siii{} profile, where $5\,800$~\AA~$\textless~\lambda~\textless~6\,800$~\AA.}\tabularnewline
\label{tab:snid_05hk_si}\tabularnewline
\resizebox{\columnwidth}{!}{
\begin{tabular}{cccccccc}\hline
\hline
Model & \multicolumn{2}{c}{2005 Nov. 02} & \multicolumn{2}{c}{2005 Nov. 06} & \multicolumn{2}{c}{2005 Nov. 14} & Mean \tabularnewline
(km~s$^{-1}$) &  & & & & &  & $r$lap \tabularnewline
\hline
\hline
& $r$lap & $z$ & $r$lap & $z$ & $r$lap & $z$ & \tabularnewline
\hline
$\phn\phn$138 & 0.99  & -0.008(0.018) & 1.30  & -0.009(0.000) & 0.54  & -0.009(0.000)   & 0.94   \tabularnewline
$\phn\phn$277 & 0.59  & -0.008(0.020) & 1.26  & -0.009(0.000) & 0.55  & -0.009(0.000)   & 0.80   \tabularnewline
$\phn\phn$554 & 0.83  & -0.008(0.019) & 1.25  & -0.009(0.000) & 0.48  & -0.009(0.000)   & 0.85   \tabularnewline
$\phn\phn$830 & 0.77  & -0.008(0.018) & 1.02  & -0.009(0.000) & 0.32  & -0.009(0.000)   & 0.70   \tabularnewline
$\phn$1\,522  & 0.40  & -0.007(0.021) & 0.49  & -0.009(0.000) & 0.92  & 0.019(0.015)   &  0.60  \tabularnewline
$\phn$2\,491  & 0.64  & 0.025(0.023)  & 1.11  & 0.020(0.013)  & 1.44  & 0.021(0.013)  &   1.06 \tabularnewline
$\phn$4\,429  & 0.96  & 0.024(0.018)  & 1.72  & 0.021(0.011)  & 1.91  & 0.021(0.011)  &   1.53 \tabularnewline
$\phn$7\,889  & 1.11  & 0.023(0.016)  & 1.96  & 0.021(0.010)  & 2.24  & 0.022(0.010)  &   1.77 \tabularnewline
13\,978       & 1.25  & 0.023(0.015)  & 1.92  & 0.021(0.010)  & 2.47  & 0.021(0.010)  &   1.88 \tabularnewline
24\,774       & 1.35  & 0.023(0.015)  & 1.88  & 0.021(0.011)  & 2.71  & 0.021(0.009)  &   1.98 \tabularnewline
Uniform       & 1.31  & 0.023(0.015)  & 1.85  & 0.021(0.011)  & 2.79  & 0.021(0.009)  &   1.98 \tabularnewline
N5def         & 1.24  & 0.022(0.015)  & 1.33  & 0.020(0.013)  & 2.09  & 0.021(0.011)  &   1.55 \tabularnewline
\hline
\hline
\end{tabular}
}
\end{table}

\begin{figure*}
\centering
\includegraphics[width=0.94\textwidth]{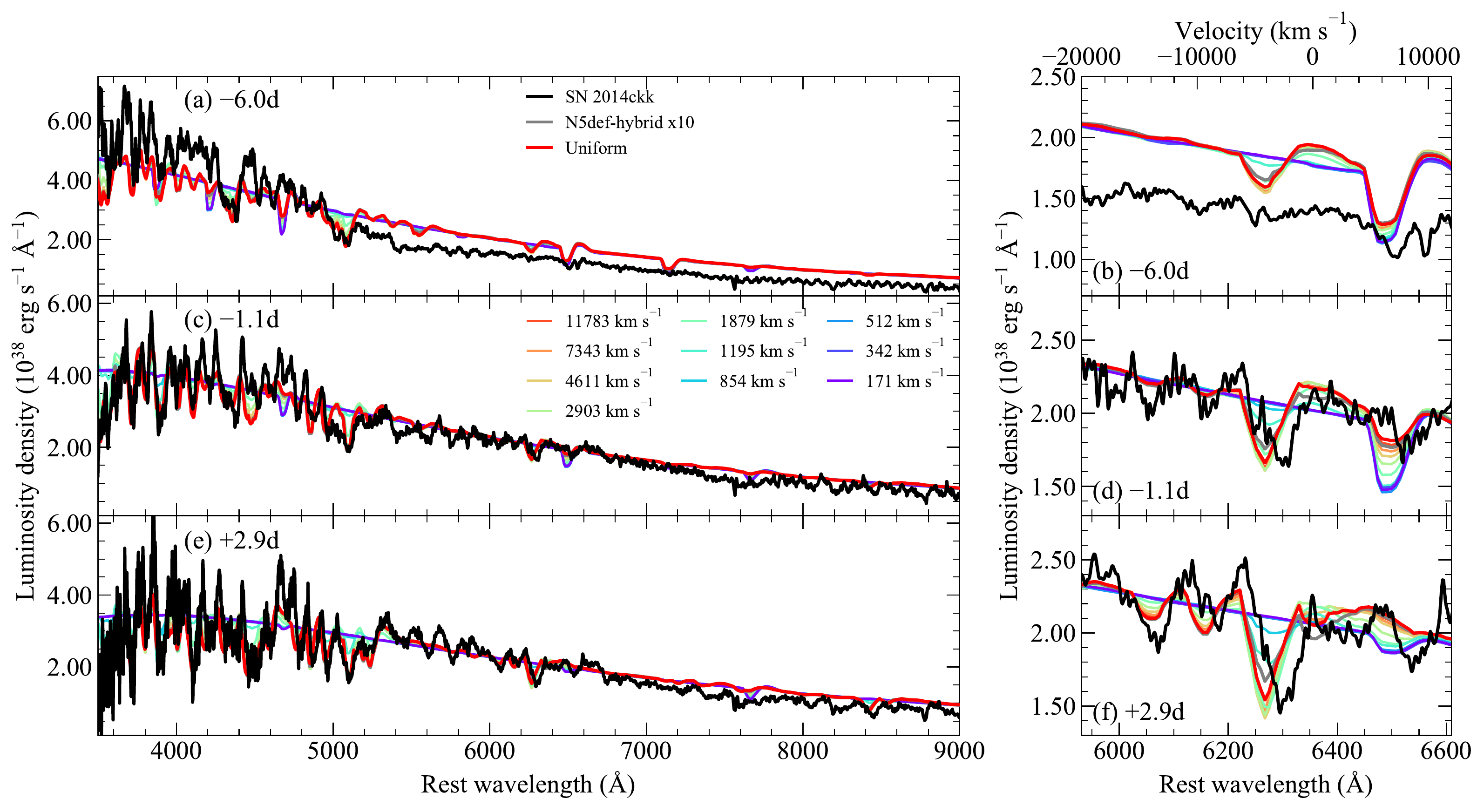}
\caption{As in Fig.~\ref{fig:12z_comp} for the case of SN~2014ck and the N5def-hybrid explosion model with an increased density \citep{kromer-15}.
}
\label{fig:14ck_comp}
\centering
\end{figure*}

\begin{table}
\centering
\caption{SNID $r$lap correlation coefficients for SN~2014ck.}\tabularnewline
\label{tab:snid_14ck}\tabularnewline
\resizebox{\columnwidth}{!}{
\begin{tabular}{rccccccc}\hline
\hline
Model & \multicolumn{2}{c}{2014 Jul. 01} & \multicolumn{2}{c}{2014 Jul. 06} & \multicolumn{2}{c}{2014 Jul. 10} & Mean \tabularnewline
(km~s$^{-1}$) &  & & & & &  & $r$lap \tabularnewline
\hline
\hline
& $r$lap & $z$ & $r$lap & $z$ & $r$lap & $z$ & \tabularnewline
\hline
$\phn\phn$171           & 1.19  & 0.957(0.010)  & 1.09  & 1.200(0.011)  & $\phn$2.32  & 1.155(0.006)   & 1.53   \tabularnewline
$\phn\phn$342           & 1.51  & 0.043(0.010)  & 1.23  & 1.200(0.000)  & $\phn$2.23  & 1.156(0.006)   & 1.66   \tabularnewline
$\phn\phn$512           & 1.35  & 0.043(0.010)  & 1.24  & 1.043(0.010)  & $\phn$2.74  & 1.047(0.006)   & 1.78   \tabularnewline
$\phn\phn$854           & 1.47  & 0.043(0.010)  & 1.34  & 0.007(0.008)  & $\phn$2.75  & 0.007(0.006)   & 1.85   \tabularnewline
$\phn$1\,195            & 1.34  & 0.958(0.010)  & 3.60  & 0.007(0.005)  & $\phn$4.94  & 0.007(0.004)   & 3.29   \tabularnewline
$\phn$1\,879            & 2.58  & 0.007(0.006)  & 6.71  & 0.007(0.003)  & $\phn$8.42  & 0.008(0.002)   & 5.90   \tabularnewline
$\phn$2\,903            & 3.99  & 0.008(0.005)  & 8.37  & 0.007(0.002)  & 13.38  & 0.008(0.001)  & 8.58   \tabularnewline
$\phn$4\,611            & 4.76  & 0.008(0.004)  & 9.06  & 0.007(0.002)  & 13.79  & 0.008(0.001)  & 9.20   \tabularnewline
7\,343                  & 4.82  & 0.008(0.004)  & 9.22  & 0.007(0.002)  & 13.42  & 0.008(0.002)  & 9.15   \tabularnewline
11\,783                 & 5.11  & 0.008(0.004)  & 9.22  & 0.007(0.002)  & 13.41  & 0.008(0.002)  & 9.25   \tabularnewline
Uniform                 & 5.41  & 0.008(0.004)  & 8.61  & 0.007(0.002)  & 12.51  & 0.008(0.002)  & 8.84   \tabularnewline
N5def-hybrid $\times$10 & 5.04  & 0.008(0.005)  & 8.39 & 0.007(0.002)   & 12.55  & 0.008(0.002)  & 8.66   \tabularnewline
\hline
\hline
\end{tabular}
}
\end{table}

\begin{table}
\centering
\caption{SNID $r$lap correlation coefficients for SN~2014ck surrounding the \siii{} profile, where $5\,800$~\AA~$\textless~\lambda~\textless~6\,800$~\AA.}\tabularnewline
\label{tab:snid_14ck_si}\tabularnewline
\resizebox{\columnwidth}{!}{
\begin{tabular}{rccccccc}\hline
\hline
Model & \multicolumn{2}{c}{2014 Jul. 01} & \multicolumn{2}{c}{2014 Jul. 06} & \multicolumn{2}{c}{2014 Jul. 10} & Mean \tabularnewline
(km~s$^{-1}$) &  & & & & &  & $r$lap \tabularnewline
\hline
\hline
& $r$lap & $z$ & $r$lap & $z$ & $r$lap & $z$ & \tabularnewline
\hline
$\phn\phn$171           & 0.75  & 0.007(0.013)  & 0.03  & 0.011(0.008)  & 0.05  & 0.036(0.018)  & 0.28   \tabularnewline
$\phn\phn$342           & 0.79  & 0.007(0.013)  & 0.04  & 0.012(0.009)  & 0.10  & 0.038(0.023)  & 0.31   \tabularnewline
$\phn\phn$512           & 0.74  & 0.007(0.013)  & 0.06  & 0.011(0.011)  & 0.03  & 0.062(0.029)  & 0.28   \tabularnewline
$\phn\phn$854           & 0.77  & 0.007(0.007)  & 0.19  & 0.009(0.014)  & 1.02  & 0.008(0.010)  & 0.66   \tabularnewline
$\phn$1\,195            & 0.96  & 0.007(0.011)  & 0.71  & 0.009(0.011)  & 1.82  & 0.009(0.007)  & 1.16   \tabularnewline
$\phn$1\,879            & 1.78  & 0.007(0.007)  & 1.69  & 0.009(0.008)  & 2.14  & 0.009(0.007) &  1.87  \tabularnewline
$\phn$2\,903            & 2.14  & 0.007(0.007)  & 2.14  & 0.009(0.007)  & 2.16  & 0.010(0.006) &  2.15  \tabularnewline
$\phn$4\,611            & 2.16  & 0.007(0.007)  & 2.32  & 0.009(0.006)  & 2.59  & 0.010(0.005) &  2.36  \tabularnewline
7\,343                  & 2.29  & 0.007(0.007)  & 2.83  & 0.009(0.005)  & 2.80  & 0.010(0.005) &  2.64  \tabularnewline
11\,783                 & 2.22  & 0.007(0.007)  & 2.86  & 0.009(0.005)  & 3.20  & 0.010(0.005) &  2.76  \tabularnewline
Uniform                 & 2.28  & 0.007(0.007)  & 2.86  & 0.009(0.005)  & 3.05  & 0.009(0.005) &  2.73  \tabularnewline
N5def-hybrid $\times$10 & 2.04  & 0.007(0.007)  & 2.74  & 0.009(0.005)  & 3.04  & 0.009(0.005) &  2.61  \tabularnewline
\hline
\hline
\end{tabular}
}
\end{table}


\bsp	
\label{lastpage}
\end{document}